\documentclass[preprintnumbers,amsmath,amssymb]{revtex4}
\usepackage[dvips]{graphicx}

\def\rf#1{(\ref{eq:#1})}
\def\lab#1{\label{eq:#1}}

\def\br{\begin{eqnarray}}
\def\er{\end{eqnarray}}
\def\be{\begin{equation}}
\def\ee{\end{equation}}
\def\({\left(}
\def\){\right)}

\def\rlx{\relax\leavevmode}
\def\IR{\rlx\hbox{\rm I\kern-.18em R}}
\def\vp{\varphi}
\def\ve{\varepsilon}
\def\u2{\mid u\mid^2}
\def\s{\sigma}
\def\pr{\prime}

\def\IZ{\rlx\hbox{\sf Z\kern-.4em Z}}
\def\IR{\rlx\hbox{\rm I\kern-.18em R}}
\def\IC{\rlx\hbox{\,$\inbar\kern-.3em{\rm C}$}}

\begin{document}

\title{Static Hopfions in the extended Skyrme-Faddeev model}
\author{L. A. Ferreira$^a$}
\email{laf@ifsc.usp.br}
\author{Nobuyuki Sawado$^{a,b}$}
\email{sawado@ph.noda.tus.ac.jp}
\author{Kouichi Toda$^{a,c}$}
\email{kouichi@yukawa.kyoto-u.ac.jp}
\affiliation{$^a$Instituto de F\'isica de S\~ao Carlos; IFSC/USP, Universidade
  de S\~ao Paulo - USP,  
Caixa Postal 369, CEP 13560-970, S\~ao Carlos-SP, Brazil\\
$^b$Department of Physics, Tokyo University of Science, Noda,
  Chiba 278-8510, Japan\\
$^c$Department of Mathematical Physics,
Toyama Prefectural University, 
Kurokawa 5180, Imizu, Toyama, 
939-0398, Japan}

\date{\today}

\begin{abstract}
We construct static soliton solutions with non-zero Hopf topological
charges to a theory which is an extension of the Skyrme-Faddeev model
by the addition of a further quartic term in derivatives. We use an
axially symmetric ansatz  based on toroidal coordinates,  and solve
the resulting two coupled non-linear partial
differential equations in two variables by a successive
over-relaxation (SOR) method. We construct numerical solutions with Hopf charge
up to four, and calculate their analytical behavior in some limiting
cases.  The solutions present an interesting 
behavior under the changes of a special combination of the coupling
constants of the quartic terms. Their energies and sizes tend to zero
as that combination approaches a particular special value. We calculate the
equivalent of the Vakulenko and Kapitanskii energy bound for the theory and
find that it vanishes at that same special value of
the coupling constants. In addition, the model presents an integrable
sector with an infinite number of local conserved currents which
apparently are not related to symmetries of the action. In the
intersection of those two special sectors the theory possesses exact
vortex solutions (static and time dependent) which were constructed in
a previous paper by one of the authors. It is believed  that
such model  describes some aspects of the low energy
limit of the pure SU(2) Yang-Mills theory, and our results may be
important in identifying important structures in that strong coupling
regime. 

\end{abstract}

\maketitle 

\section{Introduction}

We construct static soliton solutions, carrying non-trivial Hopf
topological charges, for a field theory that has found interesting
applications in many 
areas of Physics. It is a $(3+1)$-dimensional Lorentz invariant field
theory for a triplet of scalar fields ${\vec n}$, living on the
two-sphere $S^2$, ${\vec n}^2=1$, and defined by the Lagrangian density  
\be
{\cal L} = M^2\, \partial_{\mu} {\vec n}\cdot\partial^{\mu} {\vec n}
 -\frac{1}{e^2} \, \(\partial_{\mu}{\vec n} \wedge 
\partial_{\nu}{\vec n}\)^2 + \frac{\beta}{2}\,
\left(\partial_{\mu} {\vec n}\cdot\partial^{\mu} {\vec n}\right)^2
\lab{lagrangian}
\ee
where the coupling constants $e^2$ and $\beta$ are dimensionless, and
$M$ has dimension of mass. The first two terms correspond to the
so-called  Skyrme-Faddeev (SF) model \cite{Faddeev:1975}, which was
proposed long ago following Skyrme's idea \cite{skyrme}, as the
generalization to $3+1$ dimensions of the $CP^1$ model in $2+1$
dimensions \cite{bp}. The interest in the SF model has grown
considerably in recent years since the first numerical knotted soliton
solutions were constructed
\cite{Faddeev:1996zj,Battye:1998pe,Sutcliffe:2007ui,Hietarinta:1998kt,Hietarinta:2000ci},   
as well as vortex type solutions \cite{Hietarinta:vortex,hirayama}. It
has also found applications in  Bose-Einstein
condensates~\cite{Babaev:2001zy},   
superconductors~\cite{Babaev:2001jt,Babaev:2008zd}
and in the Weinberg-Salam model~\cite{Fayzullaev:2004xa}. In addition,
it has been conjectured by Faddeev and Niemi \cite{Faddeev:1998eq}
that the SF model describes the low energy (strong coupling) regime of
the pure $SU(2)$ Yang-Mills theory.  That was based on the so-called
Cho-Faddeev-Niemi-Shabanov decomposition 
\cite{Faddeev:1998eq,Cho:1979nv,Cho:1980nx,Shabanov:1999xy}  of the
$SU(2)$ Yang-Mills field ${\vec A}_{\mu}$, where its six
physical degrees of freedom are encoded into  a
triplet of scalars ${\vec n}$ (${\vec n}^2=1$), a massless $U(1)$ gauge
field, and two real 
scalar fields. The motivation for such decomposition originates from
the fact that the triplet  ${\vec n}$ can be used as an order
parameter for a condensate of Wu-Yang monopoles, which classical
solution can then be written as ${\vec A}_0=0$, ${\vec A}_{i}
=\partial_i{\vec n}\wedge {\vec n}$, $i=1,2,3$, with ${\vec n}$
corresponding to the hedgehog configuration ${\vec n}={\vec x}/r$.  

The conjecture of Faddeev and Niemi requires non-perturbative
calculations to be proved or disproved and so the discussions in the
literature have been quite controversial. Lattice field theory
simulations have discouraged its validity \cite{wipf}, and recently
Faddeev himself has proposed some modifications for it
\cite{newfaddeev}. Gies \cite{gies} has calculated the one-loop
Wilsonian effective action for the $SU(2)$ Yang-Mills theory, using
the Cho-Faddeev-Niemi-Shabanov decomposition,  and found 
agreements with the conjecture provided the SF model is modified by
additional quartic terms in derivatives of the ${\vec n}$ field. In fact,
 Gies obtained an effective action which up to first derivatives of
 ${\vec n}$ is of the form \rf{lagrangian}. Similar results were
 obtained in \cite{forkel} for the connection between the low energy
 Yang-Mills dynamics and modifications of Skyrme-Faddeev model using
 the Gaussian approximation of the vacuum wave functional.  In this sense
 the extended 
 version of the Skyrme-Faddeev model given by the theory
 \rf{lagrangian} deserves some attention. Like the SF model it presents
 the internal $O(3)$  symmetry and admit solutions with topology given
 by the Hopf map $S^3\rightarrow S^2$. The main difference is the fact
 that \rf{lagrangian} contains quartic terms in time derivatives and
 so its canonical Hamiltonian is not positive 
 definite when hard modes of the ${\vec n}$ field are
 allowed. Indeed, that is a feature of many low energy effective
 theories.  

The theory \rf{lagrangian} has many interesting aspects which deserves
attention as we now explain. It has two sectors where the static
energy density is positive definite and one of them intersects with an 
integrable sub-sector of \rf{lagrangian} with an infinite number of
conservation laws. In a Minkowiski space-time the static Hamiltonian
associated to \rf{lagrangian} is
\be
{\cal H}_{\rm static} = 
M^2\, \partial_{i} {\vec n}\cdot\partial_{i} {\vec n}
 +\frac{1}{e^2} \, \(\partial_{i}{\vec n} \wedge 
\partial_{j}{\vec n}\)^2 - \frac{\beta}{2}\,
\left(\partial_{i} {\vec n}\cdot\partial_{i} {\vec n}\right)^2
\lab{hamiltonian1}
\ee
with $i,j=1,2,3$. Therefore, it is positive definite for $M^2>0$,
$e^2>0$ and $\beta <0$. That is the sector explored in
\cite{Gladikowski:1996mb}, and where static soliton solutions with
non-trivial Hopf topological charges were constructed. 

One can now stereographic project $S^2$ on a plane and work with a
complex scalar field $u$ related to the triplet ${\vec n}$ by 
\be
{\vec n} = \frac{1}{1+\u2}\, \(u+u^*,-i\(u-u^*\),\u2 -1\)
\lab{udef}
\ee
It then follows that 
\br
{\vec n}\cdot\(\partial_{\mu}{\vec n} \wedge 
\partial_{\nu}{\vec n}\)
&=&2i\,\frac{\(\partial_{\nu} u \partial_{\mu} u^* - 
\partial_{\mu} u\partial_{\nu} u^*\)}{\(1+\u2\)^2} 
\nonumber\\
\left(\partial_{\mu} {\vec
  n}\cdot \partial^{\mu} {\vec
  n}\right)&=&4\,\frac{\partial_{\mu}u\;\partial^{\mu}u^*}{\(1+\u2\)^2}
\lab{dn2}
\er
Therefore, in terms of the field $u$ the Lagrangian \rf{lagrangian} reads
\be
{\cal L}=
4\,M^2\,\frac{\partial_{\mu}u\;\partial^{\mu}u^*}{\(1+\u2\)^2} + 
\frac{8}{e^2}\left[ 
\frac{\(\partial_{\mu}u\)^2\(\partial_{\nu}u^*\)^2}{\(1+\u2\)^4}+
\(\beta\,e^2-1\)\,\frac{\(\partial_{\mu}u\;\partial^{\mu}u^*\)^2}{\(1+\u2\)^4}
\right]
\lab{actionu}
\ee
The same static Hamiltonian \rf{hamiltonian1} is now written as 
\be
{\cal H}_{\rm static}=
4\,M^2\,\frac{\partial_{i}u\;\partial_{i}u^*}{\(1+\u2\)^2} - 
\frac{8}{e^2}\left[ 
\frac{\(\partial_{i}u\)^2\(\partial_{j}u^*\)^2}{\(1+\u2\)^4}+
\(\beta\,e^2-1\)\,\frac{\(\partial_{i}u\;\partial_{i}u^*\)^2}{\(1+\u2\)^4}
\right]
\lab{hamiltonian2}
\ee
Therefore, it is positive definite for 
\be
M^2>0\,; \qquad  e^2<0\, ; \qquad \beta <0 \, ; \qquad 
\beta\, e^2\geq 1
\lab{nicesector}
\ee
 It is that sector that we shall
consider in this paper. We show that it possesses static soliton
solutions with non-trivial Hopf topological charge. If one compares
\rf{lagrangian} with 
  the effective Lagrangian given by eq. (14) or (21) of \cite{gies}
  one observes that, in order for them to agree, $e^2$ and $\beta$
  should indeed have the same sign. In the physical limit where the
  infrared cutoff goes to zero it follows that they both become
  positive, contrary to our assumption. However, the perturbative
  calculations in \cite{gies} are 
  not valid in that limit, and so not much can be said even about their
  relative signs. An interesting study of such sector was performed in
  \cite{Sawado:2005wa} with an additional term in the action containing
  second derivatives of the ${\vec n}$ field and corresponding to the
  extra term obtained in \cite{gies} when the hard modes of ${\vec n}$
  are also integrated out.

Another interesting aspect of the theory \rf{lagrangian} has to do
with its integrability properties. The Euler-Lagrange equations
following from \rf{actionu}, or equivalently \rf{lagrangian},  reads   
\be
\(1+\u2\)\, \partial^{\mu}{\cal K}_{\mu}-2\,u^{*}\,{\cal K}_{\mu}\,
\partial^{\mu} u=0
\lab{eqmot}
\ee
together with its complex conjugate, and where
\be
{\cal K}_{\mu}\equiv M^2\, \partial_{\mu}u 
+\frac{4}{e^2}\,\frac{ 
\left[\(\beta\,e^2-1\)\,\(\partial_{\nu}u\,\partial^{\nu}u^{*}\)\,
\partial_{\mu} u+
\(\partial_{\nu}u\partial^{\nu} u\)
\partial_{\mu}u^{*}\right]}{\(1+\u2\)^2}
\lab{kdef}
\ee
The theory \rf{lagrangian} has three conserved currents associated the
$O(3)$ internal symmetry. However, using the techniques of \cite{afs,reviewafs}
one can show that the sub-sector defined by the constraint 
\be
\partial_{\mu}u\, \partial^{\mu}u=0
\lab{constraint}
\ee
possesses an infinite number of conserved currents given by 
\be
J_{\mu}\equiv \frac{\delta G}{\delta u}\, {\cal K}^{c}_{\mu}   - 
\frac{\delta G}{\delta u^{*}}\, {{\cal K}^{c}_{\mu}}^{*}
\lab{currents}
\ee
where $G$ is any function of $u$ and $u^{*}$, but not of
their derivatives, and ${\cal K}^{c}_{\mu}$ is obtained from \rf{kdef}
by imposing \rf{constraint}, i.e.
\be
{\cal K}^{c}_{\mu}\equiv M^2\, \partial_{\mu}u+\frac{4}{e^2}\,\frac{ 
\(\beta\,e^2-1\)\,\(\partial_{\nu}u\,\partial^{\nu}u^{*}\)\,
\partial_{\mu} u}{\(1+\u2\)^2}
\lab{kcdef}
\ee
The fact that  currents \rf{currents} are conserved follows from the
identity $Im\({\cal K}_{\mu}\partial^{\mu}u^*\)=0$, the condition 
${\cal  K}^c_{\mu}\partial^{\mu}u=0$ following from \rf{constraint}, and the
equations of motion, which now read $\partial^{\mu}{\cal K}^{c}_{\mu}=0$. 

If in addition of \rf{constraint} one restricts to the sector where
the coupling constants satisfy
\be
\beta\,e^2=1
\lab{bogo3}
\ee
then the equations of motion simplify to $\partial^2\,u=0$, and the theory
becomes scaling invariant. That constitutes an integrable sub-sector
of the theory \rf{lagrangian}. Exact vortex solutions were constructed
 in \cite{vortexlaf}, using quite simple and direct methods. One can
 have multi-vortex solutions all lying in the same direction, and the
 vortices can be either static or have waves traveling along them with
 the speed of light.  

Apparently the integrable sub-sector defined by conditions
\rf{constraint} and \rf{bogo3} do not admit soliton solutions with
non-trivial Hopf topological charges. In this paper we construct such
solitons for the theory \rf{lagrangian} in the range of the coupling
constants defined in \rf{nicesector} and without imposing the
constraint \rf{constraint}. We construct static solutions
with axial symmetry by solving numerically the equations of
motion. Due to the axial symmetry we have to solve two coupled
non-linear partial differential equations in two dimensions. We choose
to use the toroidal coordinates  defined as 
\br
x^1&=&\frac{r_0}{p}\,\sqrt{z}\,\cos\varphi \nonumber\\
x^2&=&\frac{r_0}{p}\,\sqrt{z}\,\sin\varphi\qquad \qquad\qquad\qquad 
p=1-\cos\xi\,\sqrt{1-z}\nonumber\\ 
x^3&=&\frac{r_0}{p}\,\sqrt{1-z}\,\sin\xi
\lab{toroidal}
\er
where $x^i$, $i=1,2,3$, are the Cartesian coordinates in $\IR^3$, and
$\(z,\xi,\varphi\)$ are the toroidal coordinates. We have $0\leqq
z\leqq 1$, $-\pi\leqq \xi\leqq \pi$, $0\leqq \varphi\leqq 2\pi$,
and $r_0$ is a free parameter with dimension of length. Notice that
the usual toroidal coordinates have a coordinate $\eta >0$
which is related to our $z$ by  $z=\tanh^2\eta$. 

We construct solutions which are invariant under the diagonal $U(1)$
subgroup of the direct product of the $U(1)$ group of rotations on the
$x^1\,x^2$ plane and the $U(1)$ group of internal phase
transformations $u\rightarrow e^{i\,\alpha}\, u$. We then use the ansatz 
\be
u= \sqrt{\frac{1-g\(z,\xi\)}{g\(z,\xi\)}}\,
e^{i\,\Theta\(z,\xi\)+i\,n\,\varphi} 
\lab{ansatz}
\ee
with $n$ being an integer, and $0\leq g\leq 1$, $-\pi\leq \Theta\leq
\pi$. 

By rescaling the Cartesian coordinates as $x^i\rightarrow x^i/r_0$,
it is then clear that the equations of motion \rf{eqmot} will depend
only upon two dimensionless parameters, namely $\beta\,e^2\geq 1$ and 
\be
a^2 =  -e^2\,r_0^2\,M^2 >0
\lab{adef}
\ee
since we are assuming \rf{nicesector}. In addition, from
\rf{hamiltonian2} the static energy can be written as 
\be
E = \int d^3x \, {\cal H}_{\rm static} =  \frac{4\,M}{\mid e\mid}\,
\left[ E_2 +2\, \( E_4^{(1)}+\(\beta\,e^2-1\)\, E_4^{(2)}\)\right]
\lab{energyscaled}
\ee
in terms of the dimensionless quantities 
\br
E_2&=& M\,\mid e\mid\,
\int d^3x \, \frac{\partial_{x^i}u\;\partial_{x^i}u^*}{\(1+\u2\)^2}
\nonumber\\
E_4^{(1)}&=& \frac{1}{M\,\mid e\mid}\,\int d^3x \,
\frac{\mid\(\partial_{x^i}u\)^2\mid^2}{\(1+\u2\)^4} 
\lab{energycompscaled}\\
E_4^{(2)}&=& \frac{1}{M\,\mid e\mid}\,\int d^3x
\,\frac{\(\partial_{x^i}u\;\partial_{x^i}u^*\)^2}{\(1+\u2\)^4} \nonumber
\er

By replacing the ansatz \rf{ansatz} into the equations of motion
\rf{eqmot} we get two non-linear partial differential equations for
the function $g\(z\, ,\xi\)$ and $\Theta\(z\, ,\xi\)$, depending upon the
dimensionless parameters $\beta\,e^2$ and $a$. We solve those
equations numerically using a standard relaxation method
\cite{relaxation}, with appropriate boundary conditions explained
below. Derrick's argument \cite{derrick} implies that the 
contribution to the energy from the 
quadractic and quartic terms must equal, i.e. 
 $E_2 =2\, \( E_4^{(1)}+\(\beta\,e^2-1\)\, E_4^{(2)}\)$ (see
 \rf{energyscaled}) . That
fixes the size of the solution and so the value of $a$. Therefore, in our
numerical procedure we determine $a$ on each step 
of the relaxation method using Derrick's argument.   
Our main results  are the following. We find finite energy axially
symmetric solutions with Hopf charges up to four. The axial symmetry
comes from the use of the ansatz \rf{ansatz}. Therefore, our solutions
are not necessarily the ones with the lowest possible energy for a
given value of the Hopf charge. For the Skyrme-Faddeev model the
solutions with Hopf charge $1$ and $2$ do present axial symmetry
\cite{Faddeev:1996zj,Battye:1998pe,Sutcliffe:2007ui,Hietarinta:1998kt,Hietarinta:2000ci}. We 
believe the same may happen in the extended model \rf{lagrangian}. An
interesting discovery we made is that $a\rightarrow 0$ as 
$\beta\,e^2\rightarrow 1$.  That means that the solutions shrink in that
limit, and disappear for $\beta\,e^2= 1$. In addition, the energy also
vanishes in the limit $\beta\,e^2\rightarrow 1$. Since all three terms
in the energy \rf{energyscaled} are positive they each vanish in that
limit, including $E_4^{(2)}$, despite the factor $\(\beta\,e^2-1\)$
in front of it. Another interesting discovery is that $E_4^{(1)}$ is
very small compared to the other two terms. In fact, for a wide range
of the parameter $\beta\,e^2$ we have
$$
\frac{E_4^{(1)}}{E_2 +2\, \( E_4^{(1)}+\(\beta\,e^2-1\)\, E_4^{(2)}\)}
\sim 10^{-3}
$$
Notice that $E_4^{(1)}$ is a good measure of how close the solutions
are of satisfying the constraint \rf{constraint}. Therefore our 
solutions are very close of belonging to the integrable sector
possessing the infinite number of conserved currents given by
\rf{currents}. It would be very interesting to perform calculations
with a finer mesh to improve the precision of that ratio. One thing is
clear however, the closer is $\beta\,e^2$ to 
unity the better the constraint is satisfied by the
solutions. However, the solutions shrink in that limit. The Hopfions
then disappear and seem to give place to the 
vortex solutions constructed in \cite{vortexlaf}. That point certainly
deserves further studies. It would be very desirable to perform
simulations to see how the vortices of  \cite{vortexlaf} behave when
the conditions \rf{constraint} and \rf{bogo3} are relaxed, and that
fact is now under investigation.  

The paper is organized as follows: in section \ref{sec:hopf} we
calculate the Hopf topological charges for the configurations within
the ansatz \rf{ansatz}, in section \ref{sec:bound} we calculate a
bound for the static energy \rf{energyscaled} using methods similar to
those of references \cite{vakulenko,wardbound} employed in case of the
Skyrme-Faddeev model. The equations of motion in terms of the ansatz
functions $g\(z,\xi\)$ and $\Theta\(z,\xi\)$  are calculated in
section \ref{sec:eqmot}, and in section \ref{sec:limits} we analyze
the properties of the solutions in three limiting cases, at spatial
infinity, at the circle $x_1^2+x_2^2=r_0^2$, with $x_3=0$, and also at
the $x^3$-axis. The numerical methods and solutions are described in section
\ref{sec:numerical}.  

\section{The Hopf  topological charge}
\label{sec:hopf}

The static solutions of the theory \rf{lagrangian} define  maps from the three
dimensional space $\IR^3$ to the target space $S^2$.  In order to have
finite energy solutions we need the fields to go to a 
vacuum configuration at spatial infinity, and so we need ${\vec
  n}\rightarrow {\rm constant}$ for $r\rightarrow \infty$, with $r$ being
the distance to the origin of the coordinate system. Therefore, finite
energy solutions map all points at spatial infinity to a fixed point
of $S^2$. Consequently, as far as the topological properties of such
maps are concerned we can identify the points at infinity and consider
the space to be $S^3$ instead of $\IR^3$. Then the finite energy solutions
 define maps $S^3\rightarrow S^2$, and those are
classified into homotopy classes labeled by an integer $Q_H$ called the Hopf
index. Such index can be calculated through an integral formula as
follows \cite{bott,afz99}. We first consider the mapping of the three
dimensional space  $\IR^3$ into a $3$-sphere
$S_Z^3$, parametrized by two complex numbers $Z_k$, $k=1,2$, such that
$\mid Z_1\mid^2+\mid Z_2\mid^2=1$. Using the ansatz \rf{ansatz} we
choose such map to be given by
\be
\IR^3\rightarrow S_Z^3:\qquad\qquad \qquad Z_1 = \sqrt{1-g}\,
e^{i\,\Theta}\qquad\qquad Z_2= \sqrt{g}\,e^{-i\,n\,\vp}
\lab{firstmap}
\ee
We then map such $3$-sphere into the target $S^2$ by  
\be
S_Z^3\rightarrow S^2:\qquad\qquad \qquad u=\frac{Z_1}{Z_2} 
\lab{secondmap}
\ee
In fact, the mapping of the plane parametrized by the complex field
$u$ into the $S^2$, parametrized by the triplet of scalar fields ${\vec
  n}$, is given by the the stereographic 
projection defined in \rf{udef}. The Hopf index of such map is given
by \cite{bott}
\be
Q_H= \frac{1}{4\,\pi^2}\,\int d^3x\, {\vec A}\cdot\({\vec \nabla}\wedge
{\vec A}\)
\lab{hopfindex}
\ee
with
\be
{\vec A}=\frac{i}{2}\sum_{k=1}^2\left[ Z_k^*\,{\vec
    \nabla}Z_k-Z_k\,{\vec \nabla}Z_k^*\right] 
\lab{ahopfdef}
\ee
and where the integral is over the coordinates $x^i$ of the three
dimensional space $\IR^3$. Even though ${\vec A}$ cannot be written
locally in terms of $u$ or ${\vec n}$, its curvature can. Indeed, from
\rf{dn2},  \rf{secondmap} and \rf{ahopfdef} we have
\br
F_{ij}\equiv \partial_i A_j-\partial_j A_i=
\frac{1}{2}\,{\vec n}\cdot\(\partial_{i}{\vec n} \wedge 
\partial_{j}{\vec n}\)
=i\,\frac{\(\partial_{j} u \partial_{i} u^* - 
\partial_{i} u\partial_{j} u^*\)}{\(1+\u2\)^2} 
\lab{fijdef}
\er
Therefore the Hopf index \rf{hopfindex} can alternatively be written
as 
\be
Q_H= \frac{1}{8\,\pi^2}\,\int d^3x\, \varepsilon_{ijk}\,A_i\, F_{jk}
\lab{hopfindex2}
\ee

Performing the calculation using the
toroidal coordinates \rf{toroidal} we find that
\be
{\vec A}=
\frac{p}{r_0}\left[-2\,\sqrt{z\,\(1-z\)}\,\(1-g\)\,\partial_z\Theta\,
{\vec e}_z 
-\frac{\(1-g\)}{\sqrt{1-z}}\,\partial_{\xi}\Theta\,{\vec e}_{\xi}
  + n\, \frac{g}{\sqrt{z}}\,{\vec e}_{\vp}\right]
\ee
where ${\vec e}_z$, ${\vec e}_{\xi}$ and ${\vec e}_{\vp}$ are the unit
vectors in the direction of the changes of the position vector 
${\vec r}$ under variations of coordinates, i.e. ${\vec
  e}_z=\partial_z{\vec r}/\mid \partial_z{\vec r}\mid$, and so on. The
metric and volume element in toroidal coordinates \rf{toroidal} are 
\be
ds^2=\(\frac{r_0}{p}\)^2\left[\frac{dz^2}{4z\,\(1-z\)}+\(1-z\)\,d\xi^2+
z\,d\vp^2\right] \qquad\qquad\qquad
d^3x=\frac{1}{2}\,\(\frac{r_0}{p}\)^3\,dz\,d\xi\,d\vp
\ee
Therefore the Hopf index \rf{hopfindex} for the field configurations
in the ansatz \rf{ansatz} is given by 
\be
Q_H=\frac{n}{2\,\pi}\,\int_0^1dz\,\int_{-\pi}^{\pi}d\xi\,
\left[\partial_z \(g\,\partial_{\xi}\Theta\)-\partial_{\xi}\(g\partial_z
  \Theta\)\right] 
\lab{almosthopft}
\ee
We now impose the boundary conditions
\be
g\(z=0,\xi\)=0\qquad\qquad\qquad\qquad g\(z=1,\xi\)=1\qquad\qquad\qquad
{\rm for} \qquad \qquad -\pi\leq \xi\leq \pi 
\lab{bc1}
\ee
and
\be
\Theta\(z,\xi=-\pi\)=-m\,\pi
\qquad\qquad\qquad\qquad\Theta\(z,\xi=\pi\)=m\,\pi\qquad\qquad\qquad
{\rm for} \qquad \qquad 0\leq z\leq 1
\lab{bc2}
\ee
with $m$ being an integer. It then follows that
$\partial_z\Theta\mid_{\xi=-\pi}=\partial_z\Theta\mid_{\xi=\pi}=0$, and
so from \rf{almosthopft}
\be
Q_H= m\,n
\lab{hopfmn}
\ee

\section{The energy bound}
\label{sec:bound}

The static energy of the theory \rf{lagrangian} in the regime
\rf{nicesector} satisfy an energy bound very similar to that one found
by Vakulenko and Kapitanskii \cite{vakulenko} for the Skyrme-Faddeev
model. The arguments are quite simple, and we start by 
the definition of $F_{ij}$ in  \rf{fijdef} which implies that  
\be
F_{ij}^2= 2\,
\frac{\(\partial_{i}u\,\partial_{i}u^*\)^2}{\(1+\u2\)^4} -
2\, \frac{\mid\(\partial_{i}u\)^2\mid^2}{\(1+\u2\)^4}
\ee
and so
\be
\frac{\partial_{i}u\,\partial_{i}u^*}{\(1+\u2\)^2}\geq
\sqrt{\frac{F_{ij}^2}{2}} 
\lab{nicebound}
\ee
We now rewrite the static energy density \rf{hamiltonian2} as
\be
{\cal H}_{\rm static}=
4\,M^2\,\frac{\partial_{i}u\;\partial_{i}u^*}{\(1+\u2\)^2} - 
\frac{4}{e^2}\,\(\beta\,e^2-1\) F_{ij}^2 - 8\,\beta 
\,\frac{\mid\(\partial_{i}u\)^2\mid^2}{\(1+\u2\)^4}
\lab{hamiltonian3}
\ee  
Therefore, using  \rf{nicebound} and working in the regime
\rf{nicesector} one has 
\be
{\cal H}_{\rm static}\geq 4\,M^2\, \sqrt{\frac{F_{ij}^2}{2}} 
- \frac{4}{e^2}\,\(\beta\,e^2-1\) F_{ij}^2
\lab{nicebound2}
\ee
Notice that the bounds \rf{nicebound} and \rf{nicebound2} are
saturated by those configurations satisfying the constraint
\rf{constraint}. Therefore we should expect the solutions to be driven
in the direction of satisfying that constraint, and that is what we
find in our numerical simulations. 

The static energy \rf{energyscaled} should then satisfy (assuming the
regime \rf{nicesector})
\br
E&\geq&  2^{11/4}\,\frac{M}{\mid e\mid}\,\sqrt{\beta\,e^2-1}\,
\left(\int d^3 x\, \sqrt{F_{ij}^2}\right)^{1/2}\,
\left(\int d^3 x\, F_{ij}^2\right)^{1/2}\nonumber\\
&+&\left[\frac{2\,M}{2^{1/4}}\,\left(\int d^3 x\, \sqrt{F_{ij}^2}\right)^{1/2}
-\frac{2}{\mid e\mid}\,\sqrt{\beta\,e^2-1}\,
\left(\int d^3 x\, F_{ij}^2\right)^{1/2}\right]^2
\nonumber
\er
We now use the Sobolev-type inequality \cite{vakulenko,wardbound}
\be
\left(\int d^3 x\, \sqrt{F_{ij}^2}\right)\,
\left(\int d^3 x\, F_{ij}^2\right)\geq \frac{1}{C} \, 
\(\frac{1}{8\,\pi^2}\,\int d^3x\, \varepsilon_{ijk}\,A_i\, F_{jk}\)^{3/2}
\ee
where $C$ is a universal constant. We then get the energy bound 
$$
E\geq  \frac{2^{11/4}}{C^{1/2}}\,
\frac{M}{\mid e\mid}\,\sqrt{\beta\,e^2-1}\; \; Q_H^{3/4}
$$
where we have used \rf{hopfindex2}. The value of $C$ found in 
\cite{kundurybakov,wardbound} is 
$$
\frac{1}{C}= 8\; 3^{3/4} \, \sqrt{2}\,\pi^4
$$
However, Ward in \cite{wardbound} conjectures that a better value is
\be
\frac{1}{C}= 64\,\sqrt{2}\,\pi^4
\lab{wardc}
\ee
Taking Ward's value \rf{wardc} one then gets the bound
\be
E\geq  64\,\pi^2\,
\frac{M}{\mid e\mid}\,\sqrt{\beta\,e^2-1}\; \; Q_H^{3/4}
\lab{finalbound}
\ee
Notice that one should expect a decrease in the energy of the hopfions
as $\beta\,e^2$ approaches unity, and that is exactly what we find in
our numerical calculations.  

\section{The equations of motion}
\label{sec:eqmot}

By replacing the ansatz \rf{ansatz} into the equations of motion
\rf{eqmot} we get two partial non-linear differential equations for
the functions $g\(z\,,\, \xi\)$ and $\Theta\(z\,,\, \xi\)$. They are
given by
\begin{widetext}
\begin{eqnarray}
&&\hspace{-1cm}g(1-g)\Bigl[4(z(1-z))^2 (p\partial_z R_z-R_z\partial_z
p)+z(p\partial_\xi R_\xi-R_\xi\partial_\xi p)\Bigr] \nonumber \\ 
&&\hspace{-0.5cm}-(1-2g)p\Bigl[\frac{3}{2}\Bigl(4\bigl(z(1-z)\bigr)^2R_z
\partial_z g+zR_\xi\partial_\xi g\Bigr) 
-g(1-g)\Bigl(4\bigl(z(1-z)\bigr)^2S_z\partial_z
\Theta+zS_\xi\partial_\xi \Theta+(1-z)S_\varphi n\Bigr)\Bigr]=0  
\lab{equationg}
\end{eqnarray}
and 
\begin{eqnarray}
&&\hspace{-1cm}g(1-g)\Bigl[4(z(1-z))^2 (p\partial_z S_z-S_z\partial_z p)+z(p\partial_\xi S_\xi-S_\xi\partial_\xi p)\Bigr] \nonumber \\
&&\hspace{-0.5cm}-(1-2g)p\Bigl[\frac{3}{2}\Bigl(4\bigl(z(1-z)\bigr)^2S_z\partial_z g+zS_\xi\partial_\xi g\Bigr)
+g(1-g)\Bigl(4\bigl(z(1-z)\bigr)^2R_z\partial_z \Theta+zR_\xi\partial_\xi \Theta+(1-z)R_\varphi n\Bigr)\Bigr]=0
\lab{equationt} 
\end{eqnarray}
with $p$ being defined in \rf{toroidal} and  where
\begin{eqnarray}
&&R_z:=-\frac{1}{2} \Bigl(g(1-g)z(1-z)+\alpha (v_a-v_b)+\gamma
(v_a+v_b)\Bigr) \partial_z g 
-\alpha g(1-g)v_c \partial_z\Theta \nonumber\\
&&R_\xi:=-\frac{1}{2} \Bigl(g(1-g)z(1-z)+\alpha (v_a-v_b)+\gamma
(v_a+v_b)\Bigr) \partial_\xi g 
-\alpha g(1-g)v_c \partial_\xi \Theta \nonumber\\
&&R_\varphi:=-\alpha g(1-g)v_c n \nonumber\\
&&S_z:= g(1-g)\Bigl(g(1-g)z(1-z)-\alpha (v_a-v_b)+\gamma
(v_a+v_b)\Bigr) \partial_z \Theta 
+\frac{1}{2}\alpha v_c \partial_zg \nonumber\\
&&S_\xi:=  g(1-g) \Bigl(g(1-g)z(1-z)-\alpha (v_a-v_b)+\gamma
(v_a+v_b)\Bigr) \partial_\xi \Theta 
+\frac{1}{2}\alpha v_c \partial_\xi g \nonumber\\
&&S_\varphi:=ng(1-g)\Bigl(g(1-g)z(1-z)-\alpha (v_a-v_b)+\gamma
(v_a+v_b) \Bigr) \nonumber
\end{eqnarray}
where 
\be
\alpha:=4\,\frac{p^2}{a^2} \qquad\qquad\qquad\qquad 
\gamma:=4\,\frac{p^2}{a^2}\,(\beta\, e^2-1)
\lab{alphagamma}
\ee
and
\begin{eqnarray}
&&v_a:=\frac{1}{4}\Bigl(4(z(1-z))^2(\partial_z g)^2+z(\partial_\xi
g)^2\Bigr) 
\nonumber\\
&&v_b:=(g(1-g))^2\Bigl(4(z(1-z))^2(\partial_z \Theta)^2+z(\partial_\xi
\Theta)^2+(1-z)n^2\Bigr) 
\nonumber\\ 
&&v_c:=g(1-g)\Bigl(4(z(1-z))^2(\partial_z
g)(\partial_z\Theta)+z(\partial_\xi g)(\partial_\xi \Theta)\Bigr) 
\lab{vdef} 
\end{eqnarray}
\end{widetext}
Besides the boundary conditions \rf{bc1} and \rf{bc2} leading to the
Hopf index given by \rf{hopfmn}, we impose as well the additional
boundary conditions
\br
\partial_\xi g(z,\xi)|_{\xi=-\pi}=\partial_\xi g(z,\xi)|_{\xi=\pi}&=&0,
\qquad\qquad\qquad {\rm for}\qquad\qquad 0\leq z \leq 1
\lab{bc3}\\
\partial_z \Theta(z,\xi)|_{z=0}=\partial_z
\Theta(z,\xi)|_{z=1}&=&0,
\qquad\qquad\qquad {\rm for}\qquad\quad
-\pi\leq \xi \leq \pi 
\lab{bc4}
\er
Notice that the equations of motion \rf{equationg} and \rf{equationt}
are invariant under the transformations
\be
\xi \leftrightarrow -\xi \qquad\qquad 
g\(z\, ,\, \xi\) \leftrightarrow g\(z\, ,\, -\xi\) \qquad\qquad 
\Theta\(z\, ,\, \xi\) \leftrightarrow -\Theta\(z\, ,\, -\xi\) 
\lab{symmetryeqmot}
\ee
Therefore, we choose the boundary conditions
\be
\partial_\xi g(z,\xi)|_{\xi=0}=0, \qquad\qquad\qquad \Theta(z,\xi=0)=0
\lab{bc5}
\ee
and perform our numerical calculations on the half-plane defined by
$0\leq z\leq 1$ and $0\leq \xi\leq \pi$. The functions on the other
half-plane, namely $0\leq z\leq 1$ and $-\pi\leq \xi\leq 0$, are
obtained from the symmetry \rf{symmetryeqmot}.

\section{The behavior of the solutions in limiting cases}
\label{sec:limits}

The solutions within the ansatz \rf{ansatz} have axial symmetry around
the $x^3$-axis. In addition, from \rf{udef} and \rf{ansatz} one
observes that the condition $n_3={\rm constant}$ implies $g={\rm
  constant}$. The numerical solutions satisfying the boundary
conditions \rf{bc1}, \rf{bc2}, \rf{bc3}, \rf{bc4} and \rf{bc5}, that
we find are such that for a given value of $\xi$, $g$ is a
monotonically increasing function of $z$. In addition, for a given
value of $z$, $g$ does not vary much under variations of $\xi$. It then
turns out that the surfaces in $\IR^3$ corresponding to 
constant $n_3$ have a toroidal shape. They are in fact deformations of
the toroidal surfaces obtained by fixing $z$ and varying $\xi$ and
$\vp$ (see \rf{toroidal}). It is then important to analyze
the behavior of the solutions in three regions, namely the spatial
infinity and close to the $x^3$-axis where due to the boundary
condition \rf{bc1} we have $n_3=1$,  and close to the circle on the
$x^1\,x^2$-plane of radius
$r_0$ and centered at the origin, where $n_3=-1$. We now perform such
analysis.  

\subsection{The behavior at spatial infinity}

>From \rf{toroidal} one observes that
$$
r^2\equiv x_1^2 +x_2^2+x_3^2=
r_0^2\,\frac{\(1+\sqrt{1-z}\,\cos \xi\)}{\(1-\sqrt{1-z}\,\cos \xi\)} 
$$
Therefore, in the toroidal coordinates \rf{toroidal} the limit $r\rightarrow
\infty$ corresponds to the double limit $z\rightarrow 0$ and
$\xi\rightarrow 0$. We choose to perform such double limit through the
parametrization 
\be
z\sim  \ve_1^2\, \cos^2\s \qquad \qquad \xi \sim \ve_1\, \sin \s
\qquad\qquad{\rm with}\qquad \ve_1 \rightarrow 0 \qquad 0\leq \s\leq
\frac{\pi}{2} 
\lab{zxi}
\ee
That corresponds to polar coordinates on the plane
$\(\sqrt{z},\xi\)$, and the derivatives are related by
\br
\partial_z&=& \frac{1}{2\ve^2\cos\s}\,
\(\ve\,\cos\s\,\partial_{\ve}-\sin\s\,\partial_{\s}\) 
\nonumber\\
\partial_{\xi}&=&
\sin\s\,\partial_{\ve}+\frac{\cos\s}{\ve}\,\partial_{\s}
\nonumber
\er
Taking into account the boundary conditions
\rf{bc1}, \rf{bc4} and
\rf{bc5} we then assume the fields to behave as 
\be
g\sim \ve_1^{s_1}\,f\(\s\)\qquad\qquad \qquad  \Theta \sim \ve_1^{r_1}\,h\(\s\)
\lab{behaviorfields}
\ee
where $s_1$ and $r_1$ are constants, and with the test functions
satisfying 
\be
f\(\frac{\pi}{2}\)=0\qquad \quad f^{\pr}\(0\)=0 \qquad\qquad\qquad 
h\(0\)=0\qquad\quad h^{\pr}\(\frac{\pi}{2}\)=0
\lab{bcfh}
\ee  
where primes denote derivatives w.r.t. $\s$. Replacing the expansions
\rf{behaviorfields} into the equations of motion \rf{equationg} and
\rf{equationt}, and keeping only the leading terms, i.e. those with
the smallest possible power of $\ve_1$, we get the following equations
for the test functions
\be
\(\frac{f^{\pr}}{f}\)^{\pr} +\frac{1}{2}\,\(\frac{f^{\pr}}{f}\)^{2}
-\tan \s\, \frac{f^{\pr}}{f}
+\frac{1}{2}\,s_1\(s_1-2\)-\frac{2\,n^2}{\cos^2\s}=0
\lab{eqf}
\ee
and 
\be
\frac{h^{\pr\pr}}{h}+\left[\frac{f^{\pr}}{f}-\tan\s\right]\,\frac{h^{\pr}}{h}+
r_1\(r_1+s_1-1\)=0 
\lab{eqh}
\ee
A solution of \rf{eqf} satisfying the boundary conditions \rf{bcfh} is
given by 
\be
f\sim \(\cos\s\)^{2\mid n\mid} \qquad \qquad {\rm with}\qquad \qquad
s_1=2\(\mid n\mid+1\)
\lab{fsol}
\ee
Replacing \rf{fsol} into \rf{eqh} and performing the change of
variables 
$$
x=\sin^2\s \qquad \qquad 0\leq \s\leq \frac{\pi}{2} \qquad 0\leq x\leq 1
$$
we get that the equation \rf{eqh} becomes an hypergeometric equation
of the form
\be
x\(1-x\)\, \partial_x^2 h+\frac{1}{2}\,\left[1-\(2\mid
n\mid+3\)\,x\right]\,\partial_x h+
\frac{1}{4}\,r_1\(r_1+2\mid n\mid+1\)\,h=0
\lab{eqh2}
\ee
The boundary conditions \rf{bcfh} imply that 
\be
h\mid_{x=0} =0 \qquad \qquad \partial_x h\mid_{x=1}={\rm finite}
\lab{bch2}
\ee
There are two independent solutions for equation \rf{eqh2}, namely
\br
h_1&=& F\(\mid n\mid+\frac{r_1+1}{2},-\frac{r_1}{2},\frac{1}{2},x\)
\nonumber\\
h_2&=& x^{1/2}\, 
F\(\mid n\mid+\frac{r_1}{2}+1,-\frac{r_1}{2}+\frac{1}{2},\frac{3}{2},x\)
\nonumber
\er
where $F\(\alpha ,\beta ,\gamma,x\)$ is the Gauss hypergeometric
series. One can check that such series does not converge since in both
cases we have $\alpha+\beta-\gamma = \mid n\mid \geq 1$. 
We then have to choose $r_1$ in order to truncate the series. That leads
us to the allowed values for $r_1$
$$
r_1 = 2\, l +1 \qquad\qquad \qquad l=0,1,2,\ldots
$$
Since the first term of $F$ is $1$, we see that only $h_2$ satisfies
the boundary condition \rf{bch2}. Therefore, the allowed solutions for $\Theta$
 are 
\br
\Theta &\sim& \ve_1^{2l+1}\,\sin\s\, 
F\(\mid n\mid+l+\frac{3}{2},-l,\frac{3}{2},\sin^2\s\)
\nonumber \qquad\quad
l=0,1,2,\ldots
\er
In our numerical calculations only solutions corresponding to $l=0$
appear (so $F=1$), and therefore the behavior of our solutions at spatial
infinity, i.e. $z\, ,\, \xi\sim 0$,  is given by 
\be
g\sim \ve_1^{2\(\mid n\mid+1\)}\, \(\cos\s\)^{2\mid n\mid} \sim 
z^{\mid n\mid}\(z+\xi^2\) \qquad\qquad \qquad 
\Theta \sim \ve_1\, \sin\s \sim \xi
\ee
That implies that the $u$ field at spatial infinity behaves as
\be
u\rightarrow \frac{1}{\(2\,r_0\)^{\mid n\mid+1}}\, \frac{r^{2\mid
    n\mid+1}}{\rho^{\mid n\mid}}\,e^{i\,n\,\vp} 
\ee
with $\rho^2=x_1^2+x_2^2$, and $r^2=x_1^2+x_2^2+x_3^2$. Therefore, we
indeed have ${\vec n}\rightarrow \(0,0,1\)$, as $r\rightarrow \infty$.

\subsection{The behavior around the circle $x_1^2+x_2^2=r_0^2$, and $x_3=0$}

The circle $x_1^2+x_2^2=r_0^2$, and $x_3=0$ correspond to $z=1$ and
any $\xi$ and $\vp$, and so we write
$$
\ve_2\equiv1-z\rightarrow 0
$$
We assume the following behavior for the fields
\be
g\sim 1-\ve_2^{s_2}\, F\(\xi\) \qquad \qquad \Theta \sim
H\(\xi\)-\ve_2^{r_2}\,K\(\xi\)
\lab{trialz=1}
\ee
with $s_2$ and $r_2$ being constants. 
The boundary conditions \rf{bc2}, \rf{bc3}, \rf{bc4} and \rf{bc5}
impose the following conditions on the trial functions
\br
\partial_{\xi} g\mid_{\xi=0,\pm\pi}=0 &\rightarrow &
F^{\prime}\mid_{\xi=0,\pm\pi}=0 
 \nonumber\\
\Theta\mid_{\xi=0}=0 &\rightarrow & H\mid_{\xi=0}=0 \qquad
K\mid_{\xi=0}=0
 \nonumber\\
\Theta\mid_{\xi=\pm\pi}=\pm m \,\pi &\rightarrow & H\mid_{\xi=\pm
  \pi}=\pm m \, \pi \qquad K\mid_{\xi=\pm \pi}=0
 \nonumber\\
\partial_z\Theta\mid_{z=1}=0 &\rightarrow & r_2>1
\lab{bcz=1}
\er
The last condition  plays an important role in our analysis. In
addition, we do not want the 
$z$-derivative of $g$ to diverge close to $z=1$, and so we need $s_2\geq
1$. Replacing those expansions in the equations of motion
\rf{equationg} and \rf{equationt}  and
keeping only the leading terms we get
\br
&&\frac{1}{2}\,\(\frac{F^{\prime}}{F}\)^{\prime}+
\frac{1}{4}\,\(\frac{F^{\prime}}{F}\)^{2} + s_2^2-\(H^{\prime}\)^2
\lab{eqrz=1}\\
&+&
\delta_{s_2,1}\left[\frac{1}{2}\,\frac{F^{\prime}}{F}\, G_{+}^{\prime}+
G_{+}\(\frac{1}{2}\,\(\frac{F^{\prime}}{F}\)^{\prime}+
\frac{1}{4}\,\(\frac{F^{\prime}}{F}\)^{2} +1\)-\(H^{\prime}\)^2\,G_{-}+
\alpha\(F^{\prime}\,\(H^{\prime}\)^2\)^{\prime}\right]=0
\nonumber
\er
and 
\br
\frac{H^{\prime\prime}}{H^{\prime}}+\frac{F^{\prime}}{F} 
+\delta_{s_2,1}\left[\alpha\,F^{\prime}\,\left[
1+\(H^{\prime}\)^2-\frac{1}{4}\,\(\frac{F^{\prime}}{F}\)^{2}+
\frac{F^{\prime\prime}}{F}+\frac{1}{2}\,\frac{F^{\prime}}{F}\,
\frac{H^{\prime\prime}}{H^{\prime}}\right]+
G_{-}^{\prime}+\frac{1}{2}\,\frac{F^{\prime}}{F}\(G_{+}+G_{-}\)+
\frac{H^{\prime\prime}}{H^{\prime}}\,G_{-}\right]=0
\lab{eqsz=1}
\er
where  the primes denote derivatives w.r.t. $\xi$, and where
$\delta_{s_2,1}$ is the Kronecker delta and so those terms only 
exist in the case $s_2=1$. In addition we have introduced the quantities 
$$
G_{\pm}\equiv
F\left[\gamma\(1+\frac{1}{4}\(\frac{F^{\prime}}{F}\)^2+\(H^{\prime}\)^2\)
\pm \alpha \,\(1+\frac{1}{4}\(\frac{F^{\prime}}{F}\)^2-\(H^{\prime}\)^2\)\right]
$$
 and $\alpha$ and
$\gamma$ are the values of the functions \rf{alphagamma} in the limit
$z\rightarrow 1$, i.e.
$$
\alpha \sim \frac{4}{a^2}\qquad\qquad\qquad \gamma \sim
\frac{4}{a^2}\,\(\beta\, e^2-1\)
$$
Notice that the function $K\(\xi\)$ introduced in \rf{trialz=1} is not
determined in such expansion up to leading order. 

For the numerical calculations we perform, the relevant solution for
equations \rf{eqrz=1} and \rf{eqsz=1}, satisfying the boundary
conditions \rf{bcz=1} , is  
$$
s_2= \mid m\mid \qquad\qquad F=\mbox{\rm const.} \qquad \qquad H = m \, \xi
$$
and so
$$
g \sim 1-{\rm const.}\, \(1-z\)^{\mid m\mid} \qquad\qquad\qquad 
\Theta \sim m \, \xi
$$
Therefore, the behavior of the $u$ field around the circle
$x_1^2+x_2^2=r_0^2$, and $x_3=0$ is
\be
u \rightarrow {\rm const.} \,  \(1-z\)^{\mid m\mid/2}\,
e^{i\,\(m\,\xi+n\, \vp\)} \qquad\qquad \qquad z\rightarrow 1
\ee
Therefore we indeed have ${\vec n}\rightarrow \(0,0,-1\)$ on that circle.

\subsection{The behavior of the solutions around the $x^3$-axis}

>From \rf{toroidal} we see that the $x^3$-axis corresponds to $z=0$, and any
$\xi$ and $\vp$. Therefore we denote
$$
\ve_3\equiv z\rightarrow 0
$$
We assume the following behavior for the fields
\be
g\sim \ve_3^{s_3}\, {\cal F}\(\xi\) \qquad \qquad \Theta \sim
{\cal H}\(\xi\)+\ve_3^{r_3}\,{\cal K}\(\xi\)
\lab{trial}
\ee
with $s_3$ and $r_3$ being constants. 
The boundary conditions \rf{bc2}, \rf{bc3}, \rf{bc4} and \rf{bc5}
impose the following conditions on the trial functions
\br
\partial_{\xi} g\mid_{\xi=0,\pm\pi}=0 &\rightarrow &
{\cal F}^{\prime}\mid_{\xi=0,\pm\pi}=0 
 \nonumber\\
\Theta\mid_{\xi=0}=0 &\rightarrow & {\cal H}\mid_{\xi=0}=0 \qquad
{\cal K}\mid_{\xi=0}=0
 \nonumber\\
\Theta\mid_{\xi=\pm\pi}=\pm m \,\pi &\rightarrow & {\cal H}\mid_{\xi=\pm
  \pi}=\pm m \, \pi \qquad {\cal K}\mid_{\xi=\pm \pi}=0
 \nonumber\\
\partial_z\Theta\mid_{z=0}=0 &\rightarrow & r_3>1
\nonumber
\er
The last condition will play an important role in our analysis.  In
addition, we do not want the 
$z$-derivative of $g$ to diverge close to $z=0$, and so we need $s_3\geq
1$. Replacing those expansions into the equation of motion
\rf{equationg}  and keeping only
the leading terms we get
\be
\(n^2-s_3^2\)\left[1+\delta_{s_3,1}\,{\cal F}\,\gamma\,\(s_3^2+n^2\)\right]-
\(s_3^2+n^2\)\,\delta_{s_3,1}\,{\cal F}\,\alpha\,\(s_3^2-n^2\)=0
\lab{seqz=0}
\ee
where $\alpha$ and $\gamma$ are the limiting values of the functions
\rf{alphagamma} in this expansion, i.e. 
$$
\alpha=\frac{4}{a^2}\,\(1-\cos\xi\)^2\qquad\qquad 
\gamma =\frac{4}{a^2}\,\(\beta\,e^2-1\)\,\(1-\cos\xi\)^2
$$
So, apparently the only reasonable solution for \rf{seqz=0}  is 
$$
s_3=\mid n\mid
$$
Now replacing the same expansions in the equation of motion
\rf{equationt} for the cases $s_3>r_3>1$ and $1<s_3\leq r_3$ we get
the equation 
\be
\frac{{\cal H}^{\prime\prime}}{{\cal H}^{\prime}}+\frac{{\cal
    F}^{\prime}}{{\cal F}}- 
\frac{\partial_{\xi}p}{p}=0  \qquad\qquad\qquad\qquad s_3>r_3>1
\qquad\qquad 1<s_3\leq r_3
\lab{eqfhsneq1}
\ee
and so the solution is given by $\frac{{\cal F}\,{\cal
    H}^{\prime}}{p}= {\rm const.}$, where $p$ is the liming value of
the function introduced in \rf{toroidal}, i.e.   
$$
p=1-\cos\xi
$$
For the case $1=s_3< r_3$ the equation of motion \rf{equationt} gives instead
\be
\(1+2\,\gamma\,{\cal F}\)\,\frac{{\cal H}^{\prime\prime}}{{\cal H}^{\prime}}+
\(1+4\(\alpha+\gamma\)\,{\cal F}\)\, \frac{{\cal F}^{\prime}}{{\cal F}}-
\(1-2\,\gamma\,{\cal F}\)\,\frac{\partial_{\xi}p}{p}=0  \qquad\qquad 1=s_3< r_3
\lab{eqfhs=1}
\ee
We then notice that even though the function ${\cal K}$ introduced in
\rf{trial} does not enter into the equations, the exponent $r_3$
related to it does play a role in the expansion up to leading
order. The difficulty of this case however is that the expansion up to
leading order is not enough to determine the trial functions ${\cal
  F}$ and ${\cal H}$ separately. We would have to go to the next to
leading order terms to get those functions. 

\section{The numerical analysis}
\label{sec:numerical}

 The equations
\rf{equationg} and \rf{equationt} with the boundary conditions 
\rf{bc1}, \rf{bc2}, \rf{bc3}, \rf{bc4} and \rf{bc5} can be solved numerically
by the  standard successive over-relaxation (SOR)  
method~\cite{relaxation}.
Such well known  method is a powerful tool for finding
solutions of this kind of complex set of equations.  
In order to find a solution $f$ of an elliptic equation 
${\cal L}f=\rho$ where ${\cal L}$ represents some elliptic operator
and $\rho$ is the source, we  rewrite  it as a diffusion equation, 
\begin{eqnarray}
\frac{\partial u}{\partial t}=\omega({\cal L}u-\rho)\,.
\lab{diffusion}
\end{eqnarray}
The idea is that an initial distribution $u$ relaxes to an equilibrium
solution $f$ as $t \to \infty$, and  
$\omega$ is called  the over-relaxation parameter. (Normally one
chooses $1<\omega<2$ for faster convergence).  
We apply that algorithm to our equations 
by putting \rf{equationg}, \rf{equationt} into right-hand side of the 
diffusion equations \rf{diffusion}. 
One thing we should care about is the sign of the elliptic operator. 
In order to get convergent solutions we found that we must choose the r.h.s. of
\rf{diffusion} in such a way that the second derivatives of $g$ and
$\Theta$, w.r.t. $z$ and $\xi$, appear with a positive
sign. Therefore, we have used the diffusion equations as
\be
\frac{\partial g}{\partial t}= - \omega\, {\cal A} \qquad\qquad  
\frac{\partial \Theta}{\partial t}=  \omega\, {\cal B}
\ee
where ${\cal A}$ and ${\cal B}$ stand for the expressions on the
l.h.s. of \rf{equationg} and \rf{equationt} respectively, and $\omega$
was taken to be unity in most of the simulations. 

In order to estimate the derivatives in the equations, 
we have discretized the equations using central differences on a
rectangular cubic lattice ($z$,$\xi$).  
We have investigated the cases of various mesh sizes and found a good
convergence for a mesh size $(N_z,N_\xi)=(80,80)$, i.e. we divided the
intervals $0\leq z\leq 1$ and $0\leq \xi \leq \pi$ into $80$
segments each. 

The parameter $a$ introduced in \rf{adef} is determined by using Derrick's
scaling argument. Indeed, that argument implies that the contribution
to the energy coming from the quadratic and quartic terms in
derivatives should equal, i.e. from \rf{energyscaled} one has  $E_2
=2\, \( E_4^{(1)}+\(\beta\,e^2-1\)\, E_4^{(2)}\)$. Therefore, we
choose an initial value for $a$, and on each
step of the relaxation method we determine a new value for $a$ by
imposing that condition.  By using the ansatz \rf{ansatz} one finds
that the energy \rf{energyscaled} can be written as
\br
E=\frac{4M}{|e|}\int^1_0dz \int^\pi_{-\pi}d\xi 
\Bigl[a\epsilon_2(z,\xi)+\frac{2}{a}\Bigl(\epsilon^{(1)}_4(z,\xi)+(\beta
e^2-1)\epsilon^{(2)}_4(z,\xi)\Bigr)\Bigr] 
\er
with $a$ being defined in \rf{adef} and
\br
&&\epsilon_2(z,\xi)=\pi\frac{v_a+v_b}{pg(1-g)z(1-z)}\nonumber \\
&&\epsilon_4^{(1)}(z,\xi)=\pi\frac{p((v_a-v_b)^2+v_c^2)}{(g(1-g))^2(z(1-z))^2} \\
&&\epsilon_4^{(2)}(z,\xi)=\pi\frac{p((v_a+v_b)^2}{(g(1-g))^2(z(1-z))^2}
\nonumber  
\er
where $v_a$, $v_b$ and $v_c$ are defined in \rf{vdef}. Consequently,  
the value of $a$ on each step of the simulation is calculated by the formula
\be
a^2 = 2\, \frac{\int^1_0dz \int^\pi_{-\pi}d\xi 
\,\Bigl(\epsilon^{(1)}_4(z,\xi)+(\beta
e^2-1)\epsilon^{(2)}_4(z,\xi)\Bigr)}{\int^1_0dz \int^\pi_{-\pi}d\xi 
\;\epsilon_2(z,\xi)}
\ee

We have made several simulations studying how the solutions depend
upon the integers $n$ and $m$ (introduced in the ansatz \rf{ansatz} and
boundary conditions \rf{bc2} respectively) and the parameter $\beta\,
e^2$. For the particular case of $\beta\, e^2=1.1$, we have calculated
numerical solutions with Hopf topological charge up 
to $4$. The energies as well as the size of the solutions, measured by the
parameter $a$, are given on the Table \ref{tab:energies}.  We also
compare the energies with the bound given by \rf{finalbound}. Very
probably the solutions with Hopf charges $3$ and $4$ correspond to
excited states. 
For the Skyrme-Faddeev model the minimal energy solutions with Hopf charge $Q_H>2$
exhibit quite complicated structures like, twists, linking loops and knots, than the simple
planar tori \cite{Battye:1998pe,Sutcliffe:2007ui}.
Thus, the solutions with the minimum value of energy on
those sectors very probably do not have axial symmetry. The solutions
corresponding to $\(m,n\)=\(1,1\)$ and 
$\(m,n\)=\(1,2\)$, and so Hopf charges $1$ and $2$ respectively, may
correspond to the minimum of energy. That is in fact what happens in
the Skyrme-Faddeev model where the solutions of charge $1$ and $2$ do 
present axial symmetry.  

\begin{table*}
\begin{tabular}{|c|c|c|c|c|}  
\hline                            
$Q_H$ &($m$,$n$) &~~~~~$a$~~~~~&~~~~~~~~~$E$~~~~~~~~&
$E/E_{\rm bound}$   
\\ \hline
1  & (1,1) & 0.53 & 236.2 & 1.18 \\ \hline
2  & (1,2) & 0.79 & 390.6 & 1.16  \\ 
   & (2,1) & 0.66 & 466.7 & 1.39  \\ \hline
3  & (1,3) & 1.09 & 548.7 & 1.21  \\ 
   & (3,1) & 0.80 & 715.7 & 1.57           \\ \hline
4  & (1,4) & 1.41 & 697.3 & 1.23  \\       
   & (2,2) & 0.94 & 765.7 & 1.36       \\  
   & (4,1) & 0.99 & 965.7 & 1.71          \\ \hline
\end{tabular}
\caption{The static energy $E$ given by
  \rf{energyscaled} (in units of 
  $\dfrac{M}{|e|}$) and the parameter $a$, defined in \rf{adef}, for
  $\beta e^2=1.1$.  
We also compute the ratio of the energy $E$ to the bound given in
\rf{finalbound}, i.e. $E_{\rm bound} = 64\,\pi^2\,\frac{M}{\mid
  e\mid}\,\sqrt{\beta\,e^2-1}\; \; Q_H^{3/4}$. }
\label{tab:energies} 
\end{table*}

\begin{figure*}
\hspace{-2cm}
\includegraphics[height=6cm, width=8cm]{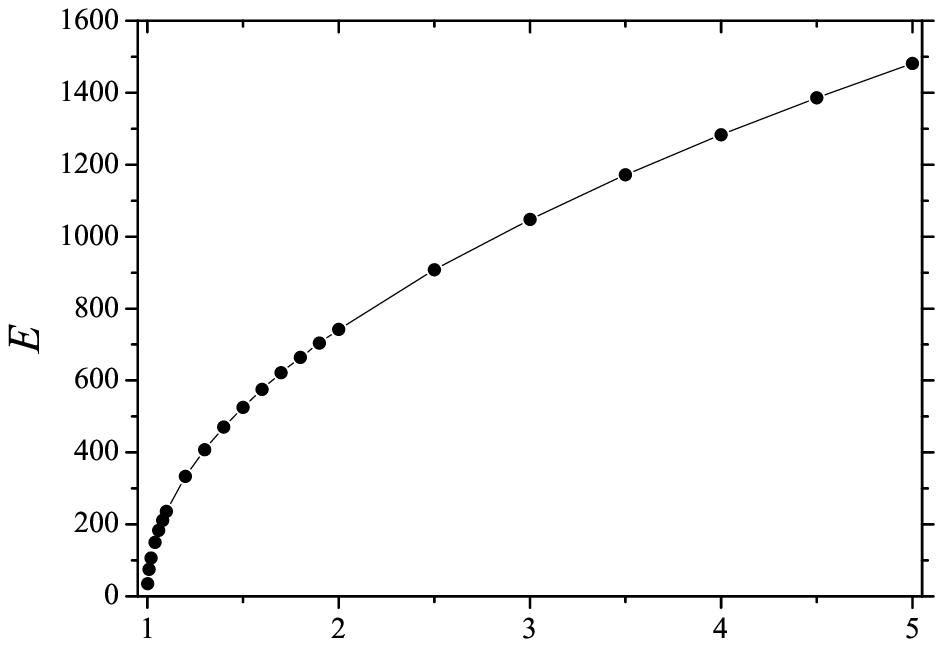}\hspace{-0.5cm}
\includegraphics[height=6cm, width=7.5cm]{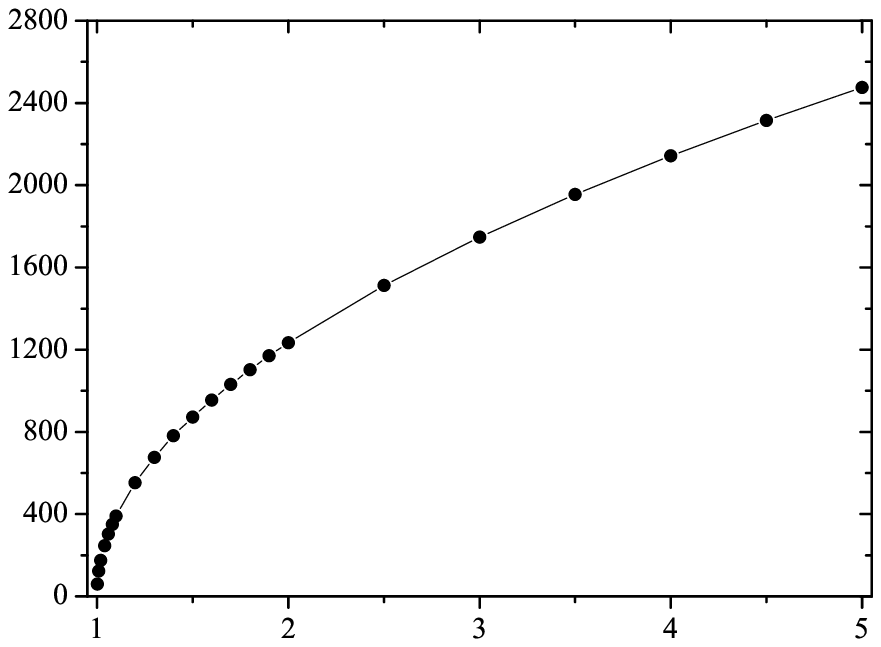}\hspace{-0.5cm}\\
\vspace{-1cm}

\hspace{-2cm}
~~\includegraphics[height=6cm, width=7.8cm]{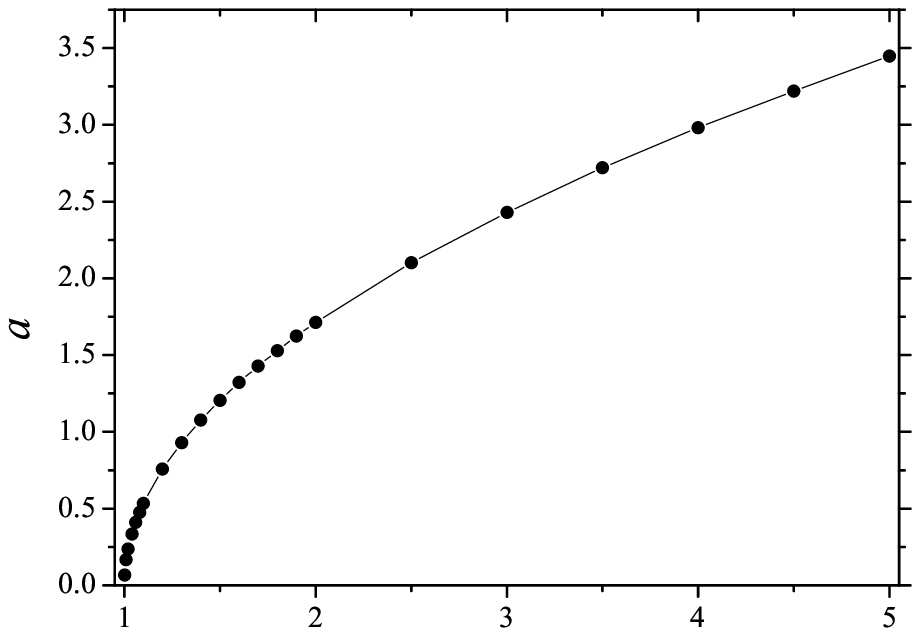}\hspace{-0.5cm}
~~\includegraphics[height=6cm, width=7.3cm]{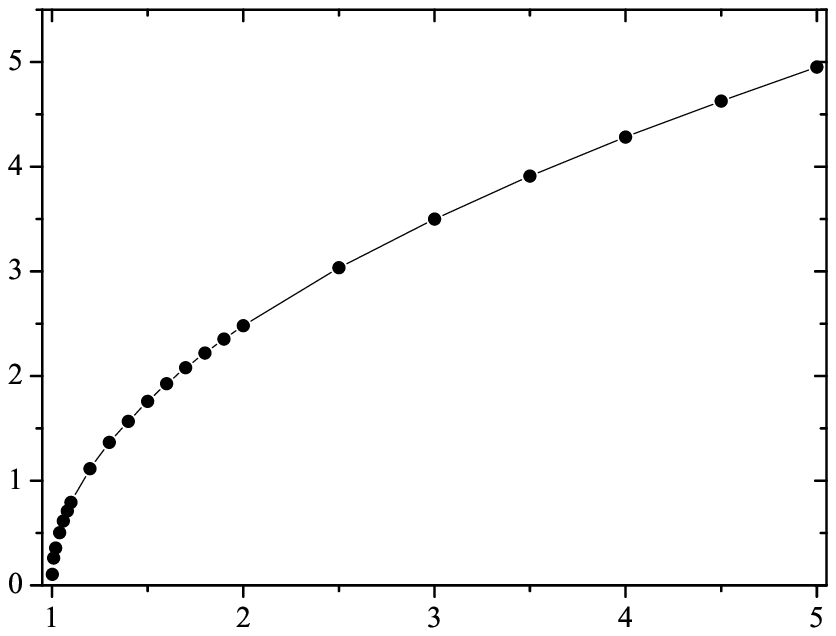}\hspace{-0.5cm}\\ 
\vspace{-1cm}

\hspace{-2cm}
\includegraphics[height=6cm, width=8cm]{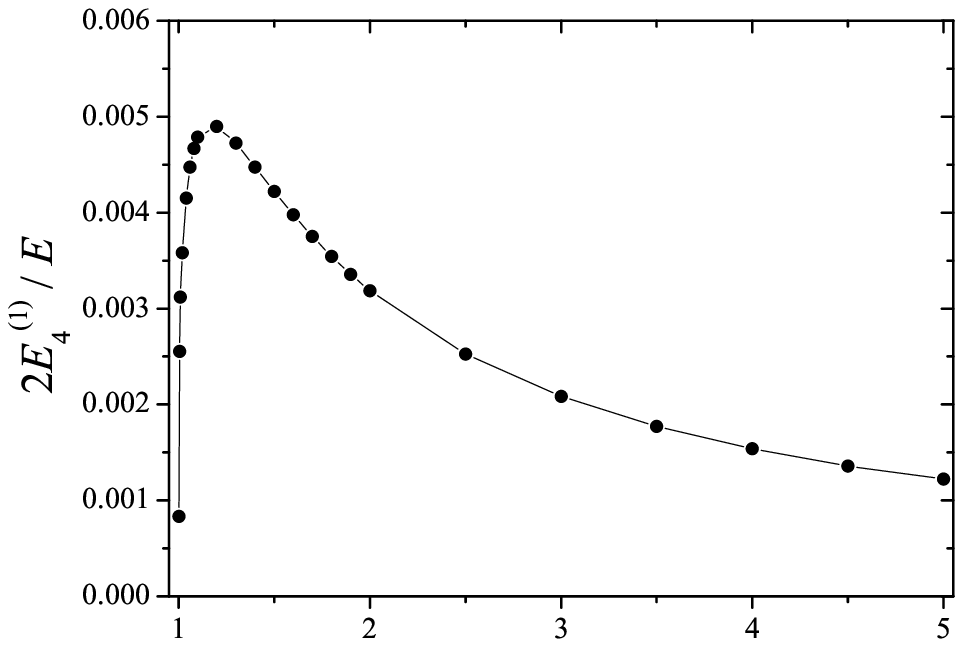}\hspace{-0.5cm}
\includegraphics[height=6cm, width=7.5cm]{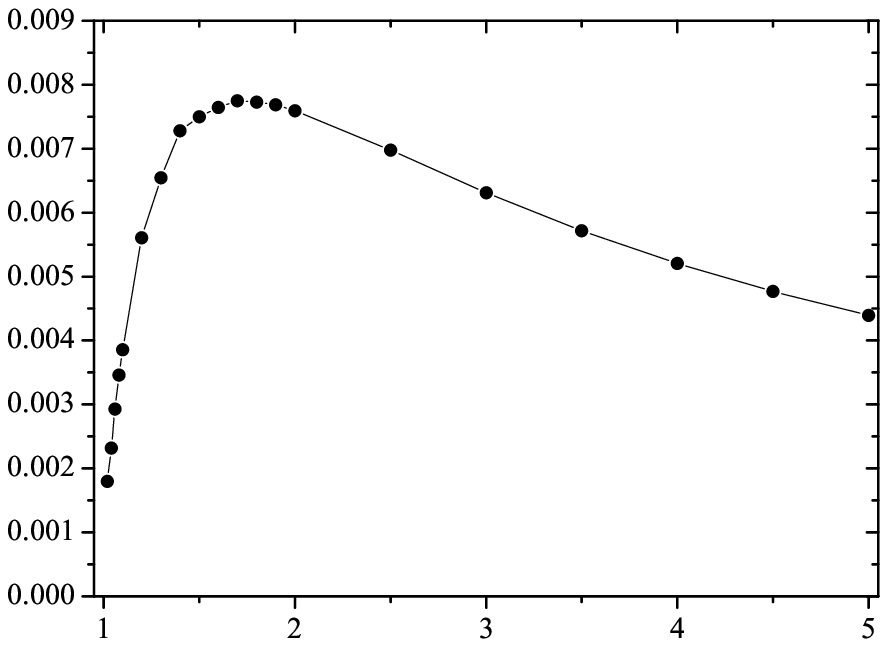}\hspace{-0.5cm}\\
\vspace{-1cm}

\hspace{-2cm}
\includegraphics[height=6cm, width=8cm]{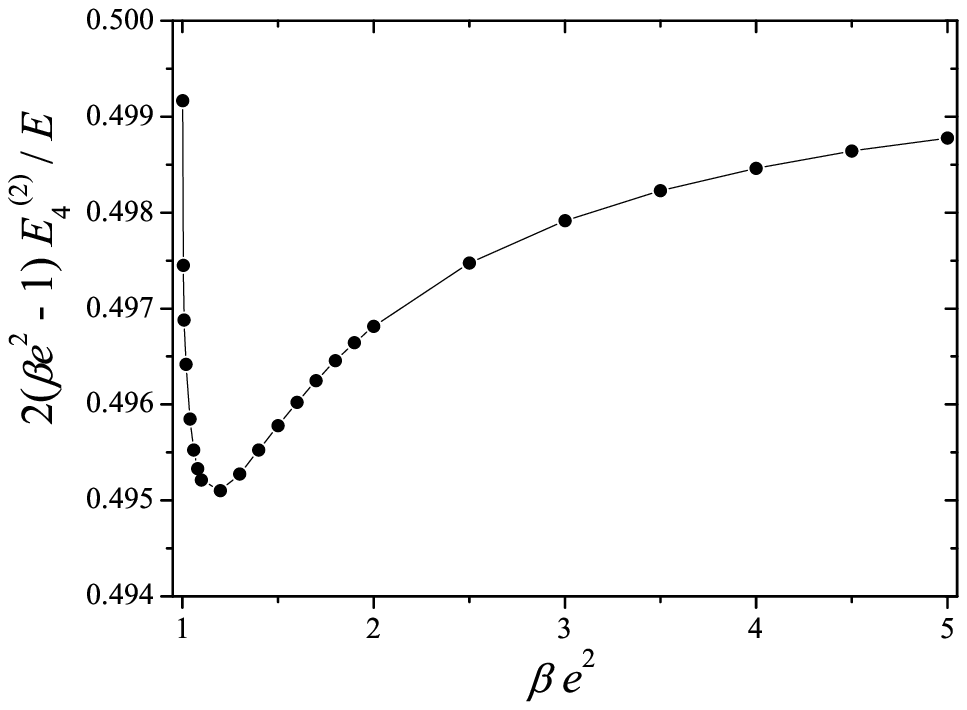}\hspace{-0.5cm}
\includegraphics[height=6cm, width=7.5cm]{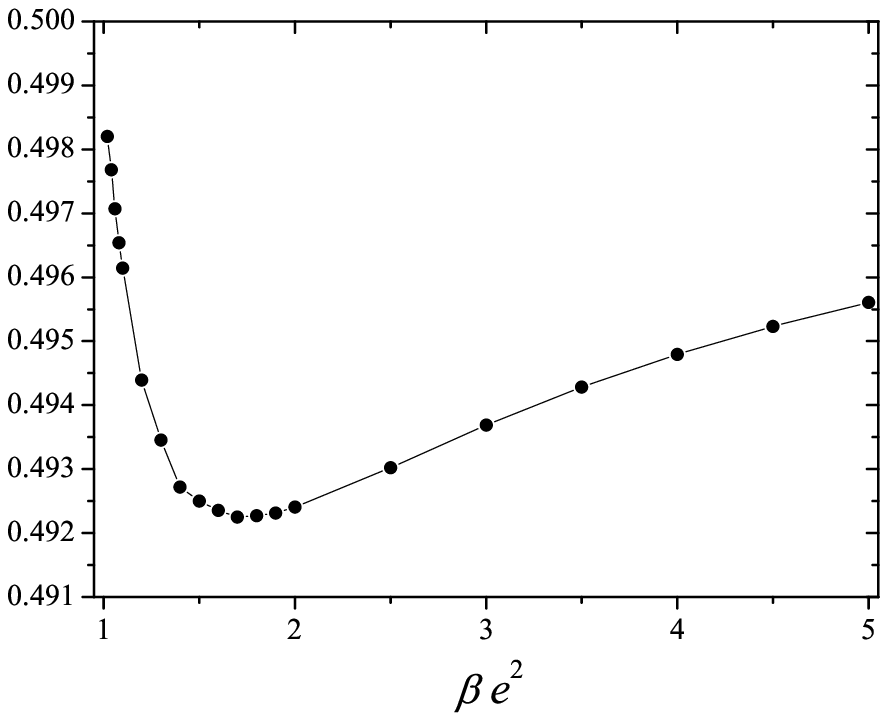}\hspace{-0.5cm}\\
\hspace{-0cm}$Q_H=1$,\; $\left[\(m,n\)=\(1,1\)\right]$
\hspace{3.5cm}$Q_H=2$,\; $\left[\(m,n\)=\(1,2\)\right]$
\caption{The plots on the l.h.s. correspond to the solution where
  $\(m,n\)=\(1,1\)$ and so Hopf charge equal to $1$, and those on
  the r.h.s to the solution where $\(m,n\)=\(1,2\)$ and so Hopf charge
  equal to $2$. On the first and second rows we plot the total energy
  $E$ (see \rf{energyscaled}) and the parameter $a$ (see \rf{adef})
  respectively, 
  against $\beta\, e^2$. On the third and fourth rows we show the
  fraction of the total energy corresponding to the two quartic terms,
  $E_4^{(1)}$ and $E_4^{(2)}$ respectively, as a function of
  $\beta\,e^2$. Notice that due to Derrick's scaling argument we must
  have $E_2=2\, \( E_4^{(1)}+\(\beta\,e^2-1\)\, E_4^{(2)}\)$, and so
  the sum of the values shown on the plots on the third and fourth
  rows   gives $0.5$.    }
\label{fig:energybetae2}
\end{figure*}

We have made a more detailed analysis of the solutions corresponding
to $\(m,n\)=\(1,1\)$ and $\(m,n\)=\(1,2\)$. We have studied how those
solutions depend upon the parameter $\beta\, e^2$, and the results are
quite interesting. The energies of those two solutions decreases as
$\beta\, e^2$ approaches unity from above, and they vanish at
$\beta\, e^2=1$. The same is true for the size of the solution. The
parameter $a$ decreases  as $\beta\, e^2\rightarrow 1$ and vanishes at
$\beta\, e^2=1$. That means that the solutions cease to exist at that
point. The results are shown in the plots of Figure
\ref{fig:energybetae2}. In addition, the contribution to the total
energy from the term 
$E_4^{(1)}$ (see \rf{energyscaled} and \rf{energycompscaled}) is very
small for the range considered for the parameter 
$\beta\, e^2$. It corresponds in fact to about $0.1\% \sim 0.5\%$ of the
total energy (see Figure \ref{fig:energybetae2}).  From
\rf{energycompscaled} we see that the term 
$E_4^{(1)}$ is a good measure of how close the solution is of
satisfying the constraint \rf{constraint}, which leads to the infinite
number of conserved currents \rf{currents}. We see that those two
solutions are very close of belonging to that integrable sector.  The
shrinking of the solutions can perhaps 
 be understood by Derrick's argument. For some reason which is not
very clear yet the dynamics of the theory keeps the quartic term of
the energy $E_4^{(1)}$ very small. We
have pointed out in section \ref{sec:bound} that configurations that
satisfy the constraint \rf{constraint} have energies closer to the
bound \rf{finalbound}, and that may be a way of understanding why
$E_4^{(1)}$ is so small. The other quartic term, namely
$E_4^{(2)}$ is multiplied by $\(\beta\, e^2-1\)$ (see
\rf{energyscaled}) and so its 
contribution vanishes as $\beta\, e^2\rightarrow 1$. Therefore, we are
left with the quadratic term $E_2$ only, which cannot be balanced by
the quartic terms anymore. Since that term scales as $E_2\rightarrow
\lambda\, E_2$, as $x^i\rightarrow \lambda\, x^i$, the solution tends
to shrink. Another fact is that the equations of motion in the sector of
the theory where $\beta\,e^2=1$, and where \rf{constraint} is
satisfied, present scale invariance (see comments below
\rf{nicesector}). Therefore, if a localized
solutions like the hopfions exists, then any re-scaling of it would
also be a solution. But that is very improbable to happen in such a
theory. Consequently, only vortex solutions like the ones 
constructed in \cite{vortexlaf} may exist in that sub-sector.     
 
\begin{figure*}
\hspace{-2cm}
\includegraphics[height=7cm, width=9cm]{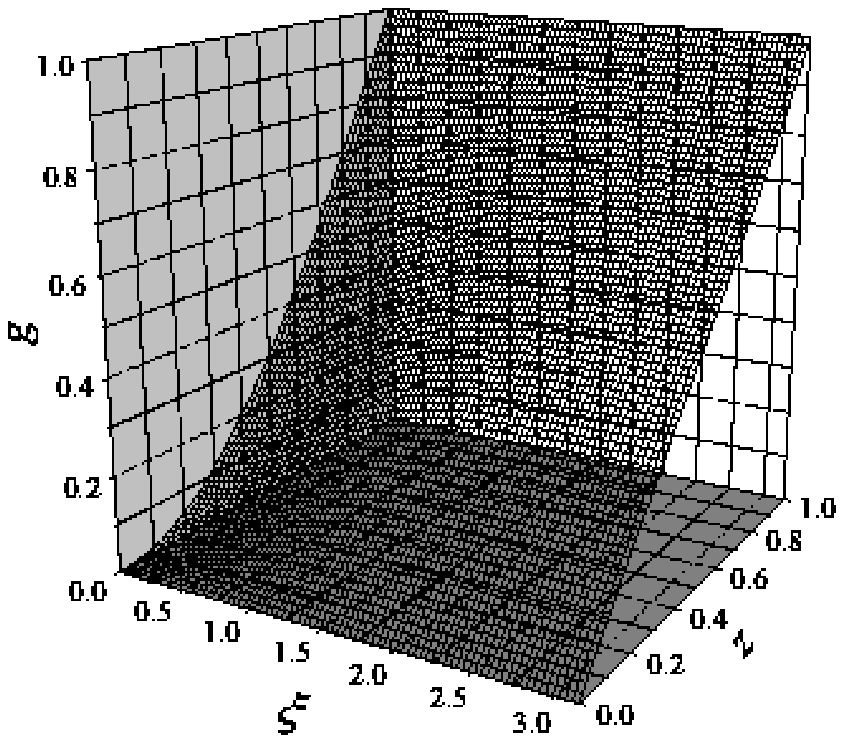}\hspace{-0.0cm}
\includegraphics[height=7cm, width=9cm]{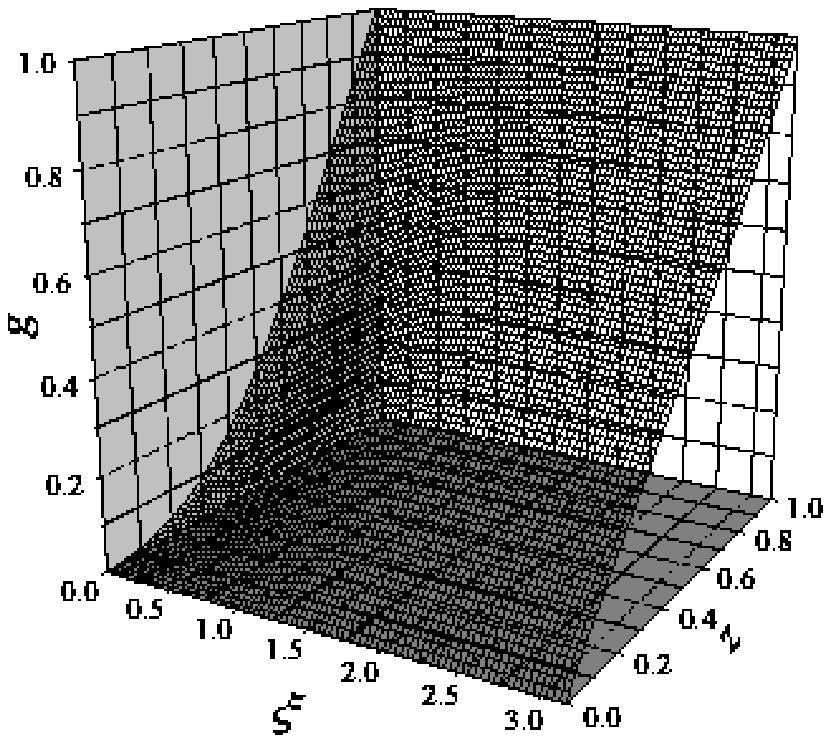}\hspace{-0.0cm}\\
\hspace{-2cm}
\includegraphics[height=7cm, width=9cm]{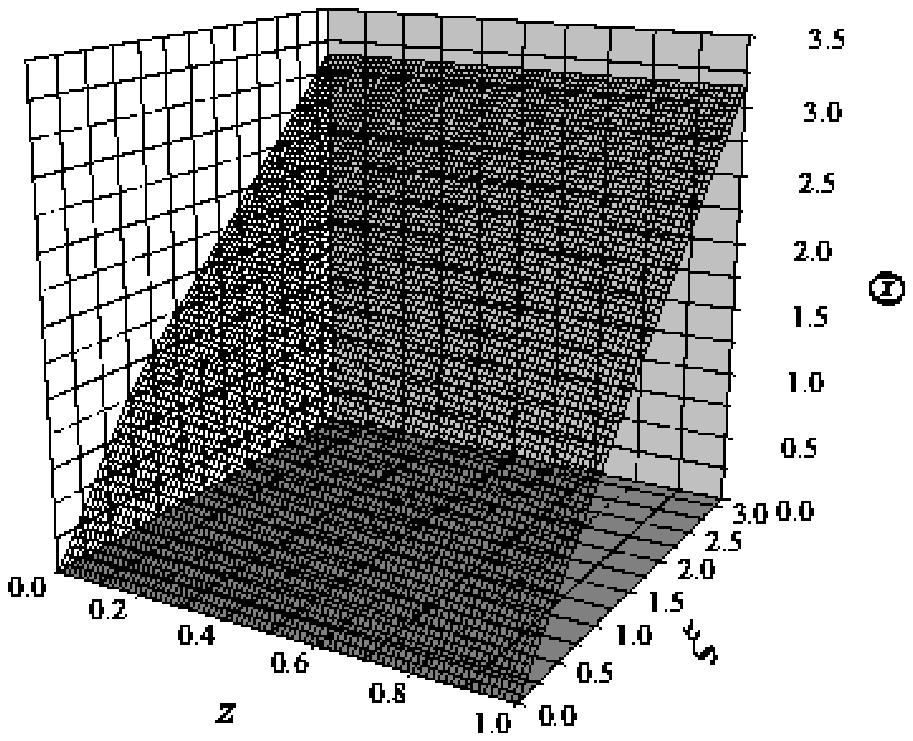}\hspace{-0.0cm}
\includegraphics[height=7cm, width=9cm]{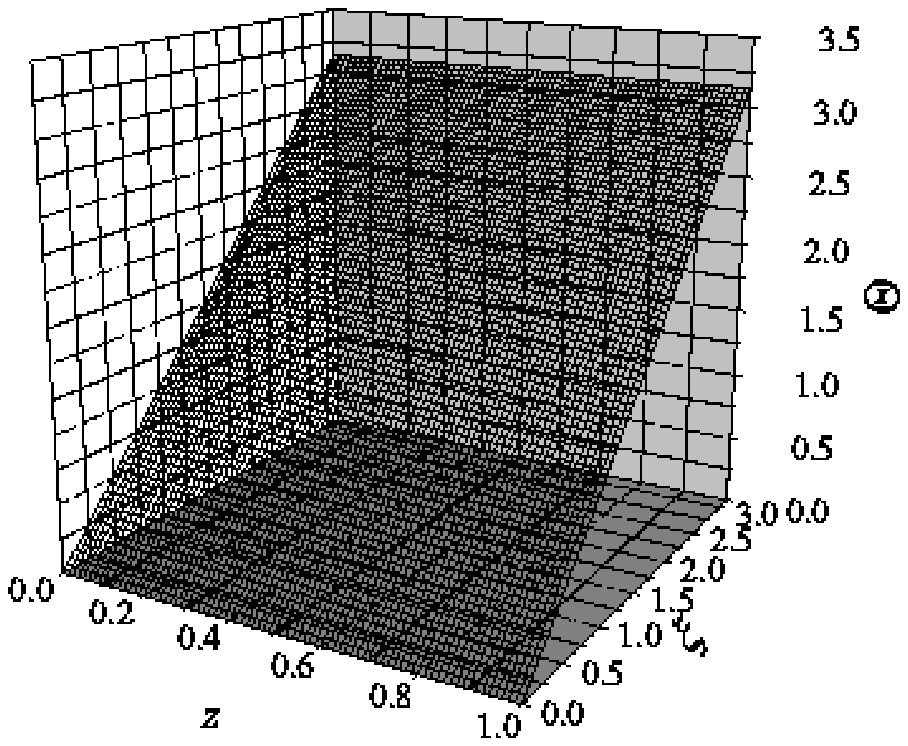}\hspace{-0.0cm}
\caption{\label{fig1}The ansatz functions  $g(z,\xi)$ and
  $\Theta(z,\xi)$ for the solution $(m,n)=(1,1)$ ($Q_H=1$)   
for $\beta e^2=1.1$ (left), and for $\beta e^2=5.0$ (right).}
\end{figure*}

\begin{figure*}
\hspace{-2cm}
\includegraphics[height=7cm, width=9cm]{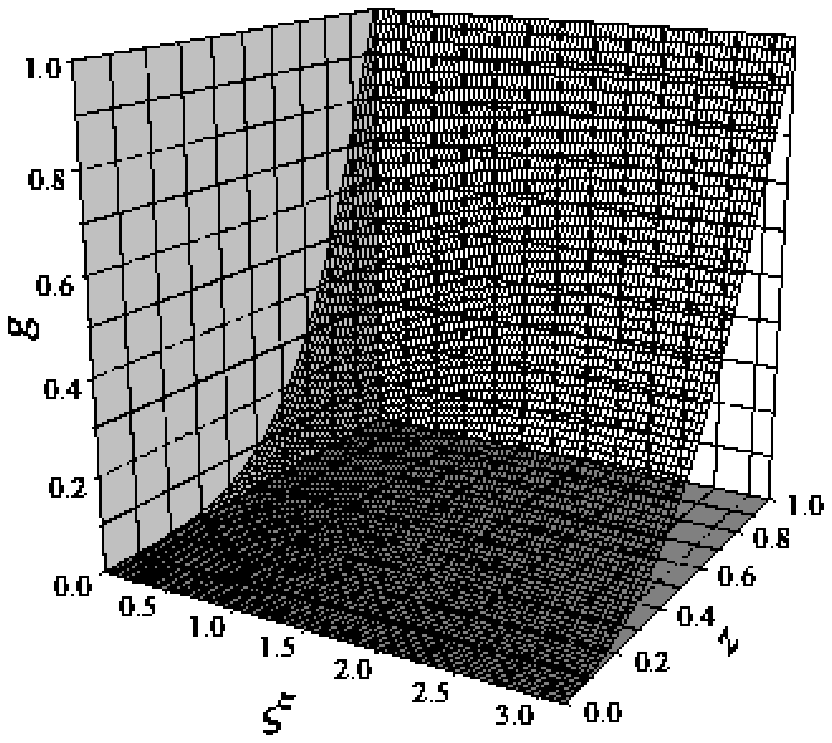}\hspace{-0.0cm}
\includegraphics[height=7cm, width=9cm]{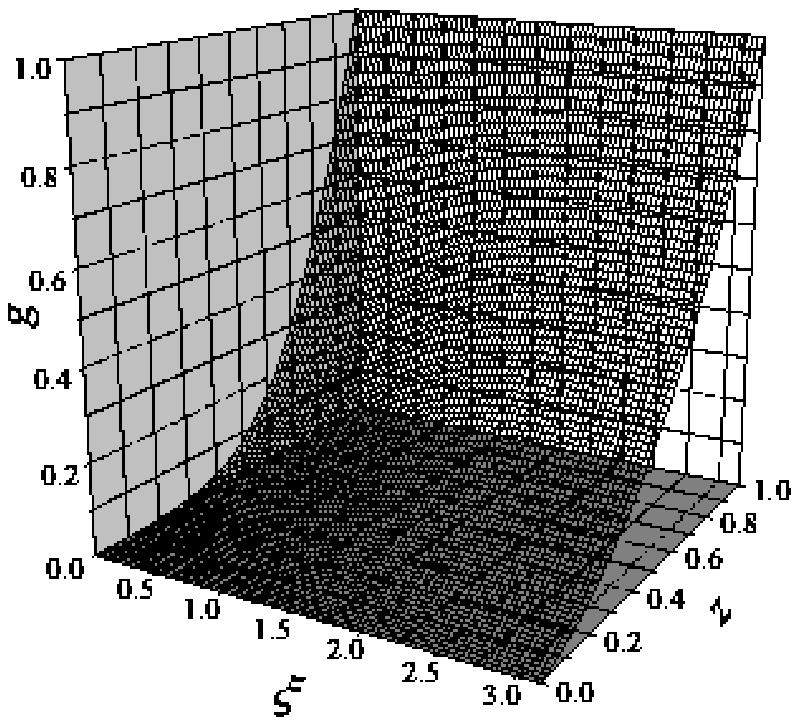}\hspace{-0.0cm}\\
\hspace{-2cm}
\includegraphics[height=7cm, width=9cm]{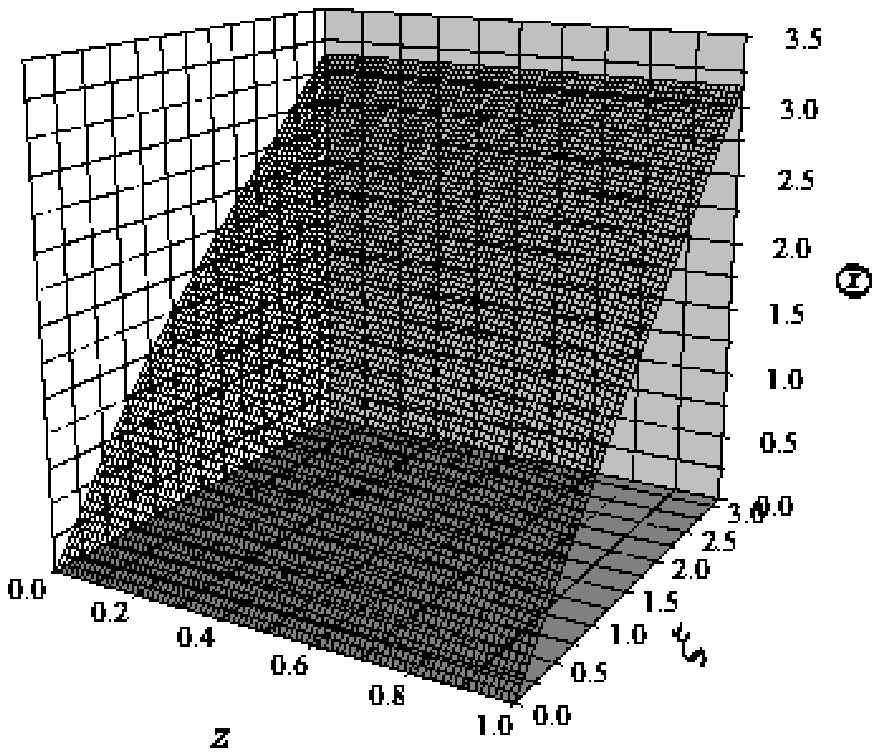}\hspace{-0.0cm}
\includegraphics[height=7cm, width=9cm]{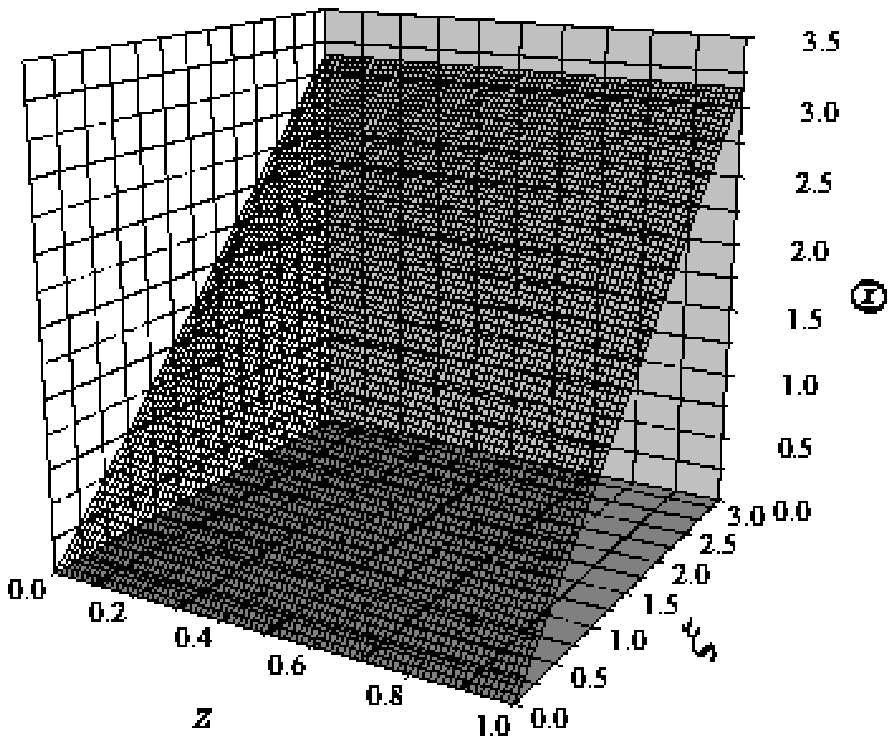}\hspace{-0.0cm}
\caption{\label{fig2}The  ansatz functions  $g(z,\xi)$ and
  $\Theta(z,\xi)$ for the solution  $(m,n)=(1,2)$ ($Q_H=2$)  
for $\beta e^2=1.1$ (left), and for $\beta e^2=5.0$ (right).}
\end{figure*}

\begin{figure*}
\hspace{-1.5cm}
\includegraphics[height=7.0cm,width=9.5cm]{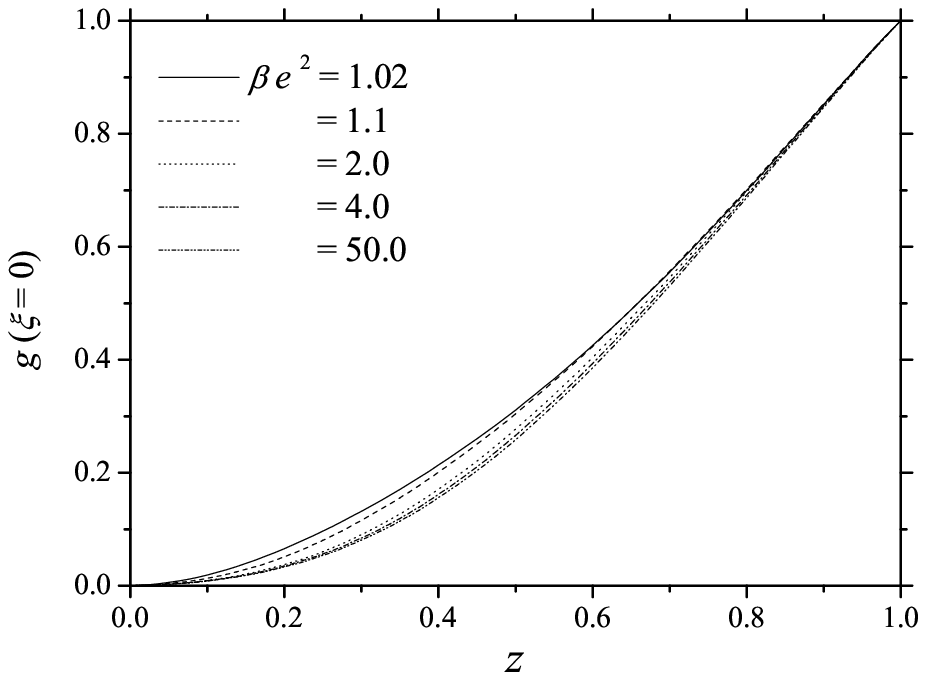}\hspace{-0.5cm} 
\includegraphics[height=7.0cm, width=9.5cm]{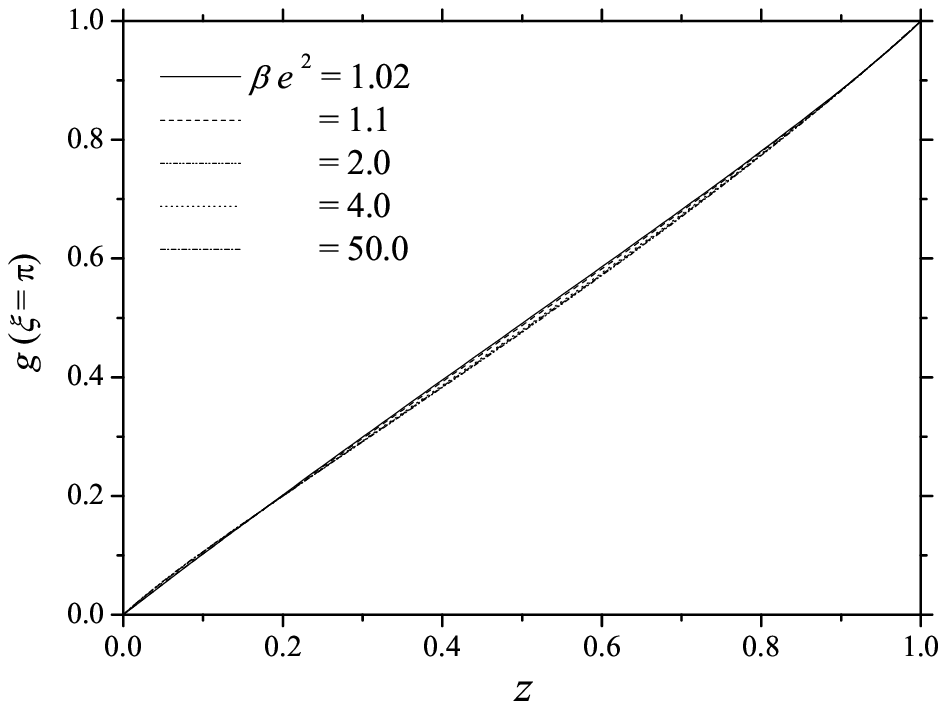}\\
{\bf (a)}~$g(z,\xi)$ at $\xi=0$ (left) and $\xi=\pi$ (right);\\  
\hspace{-1.5cm}
\includegraphics[height=7.0cm,width=9.5cm]{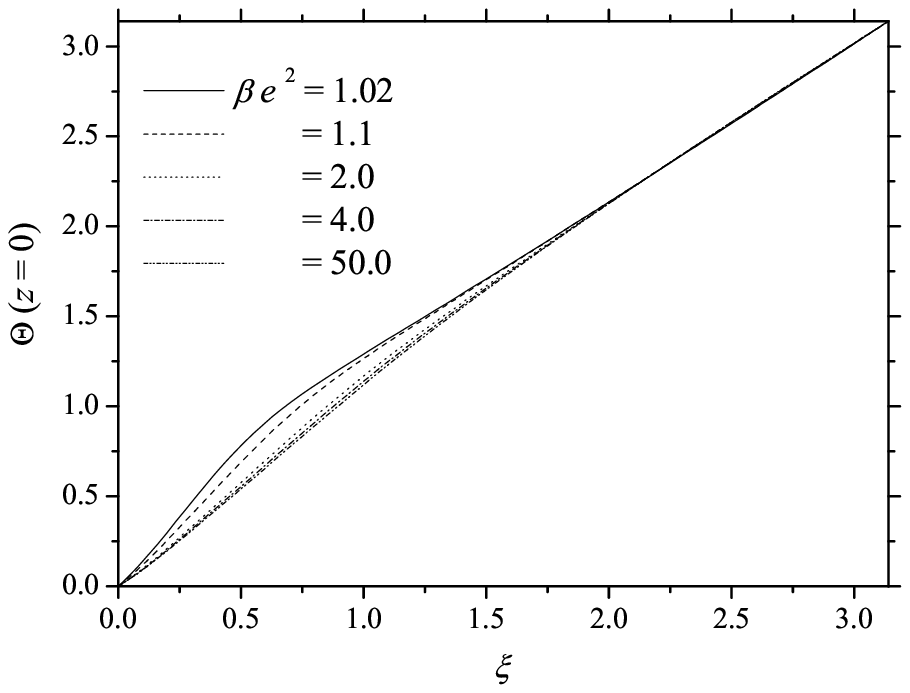}\hspace{-0.5cm} 
\includegraphics[height=7.0cm, width=9.5cm]{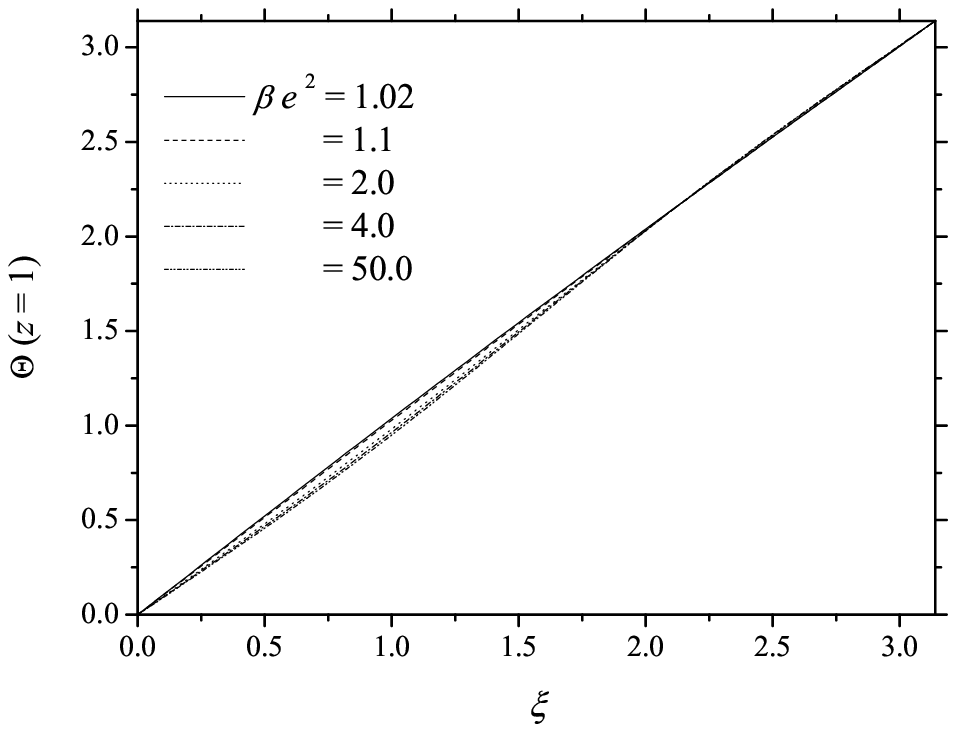}\\
{\bf (b)}~$\Theta(z,\xi)$ at $z=0$ (left) and $z=1$ (right);\\  
\caption{\label{fig3} Cuts of the ansatz functions $g(z,\xi)$ and
  $\Theta(z,\xi)$ at the boundaries for the solution $\(m,n\)=\(1,1\)$
  ($Q_H=1$) and for some values of $\beta\,e^2$ shown on the plots.}
\end{figure*}

\begin{figure*}
\hspace{-1.5cm}
\includegraphics[height=7.0cm, width=9.5cm]{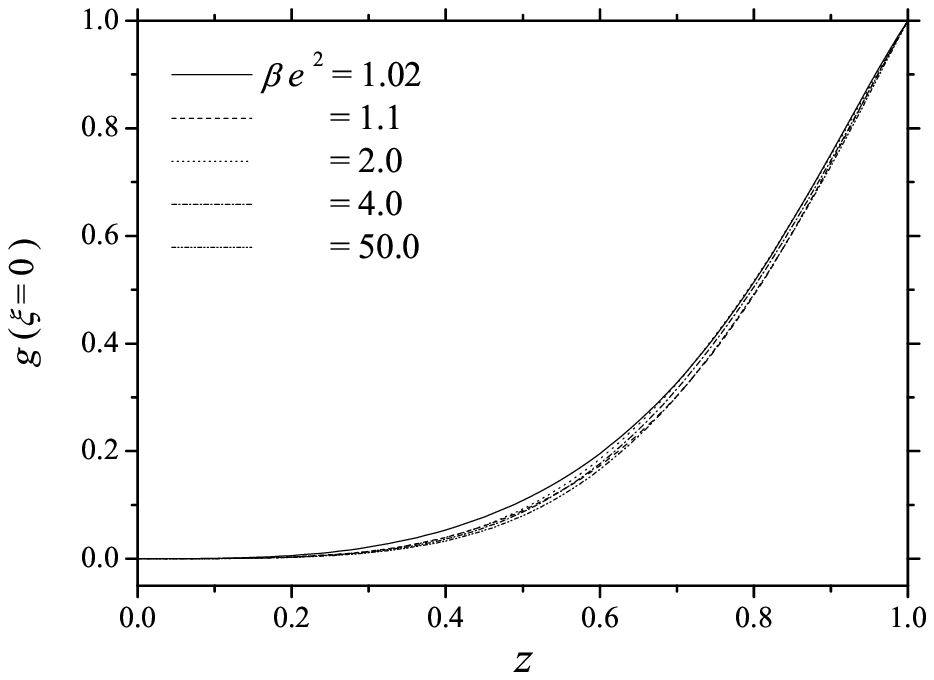}\hspace{-0.5cm} 
\includegraphics[height=7.0cm, width=9.5cm]{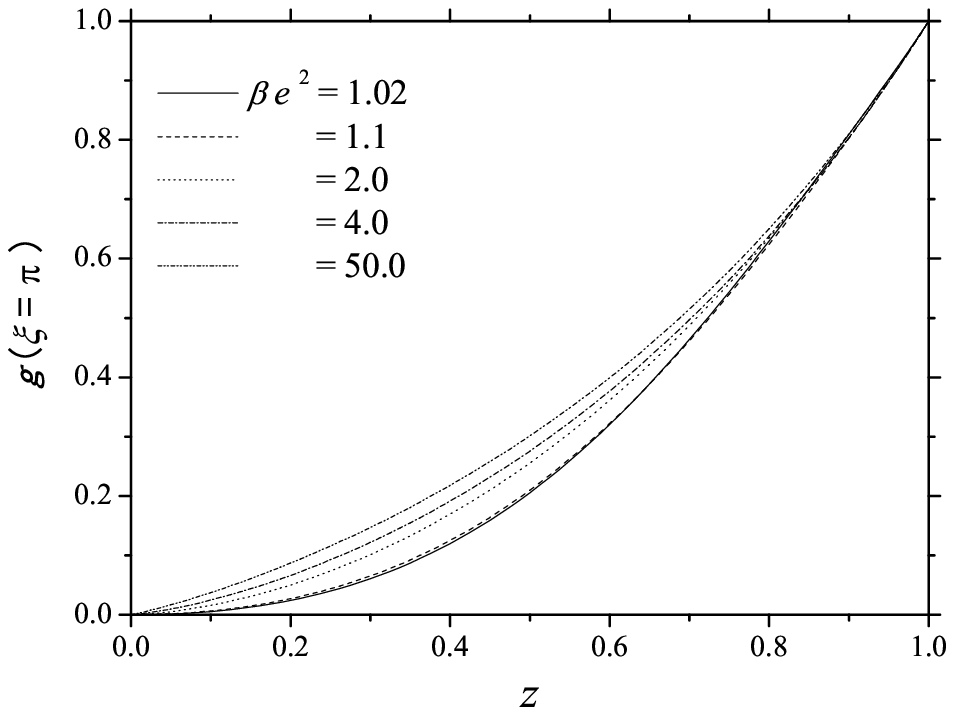}\\
{\bf (a)}~$g(z,\xi)$ at $\xi=0$ (left) and $\xi=\pi$ (right);\\  
\hspace{-1.5cm}
\includegraphics[height=7.0cm, width=9.5cm]{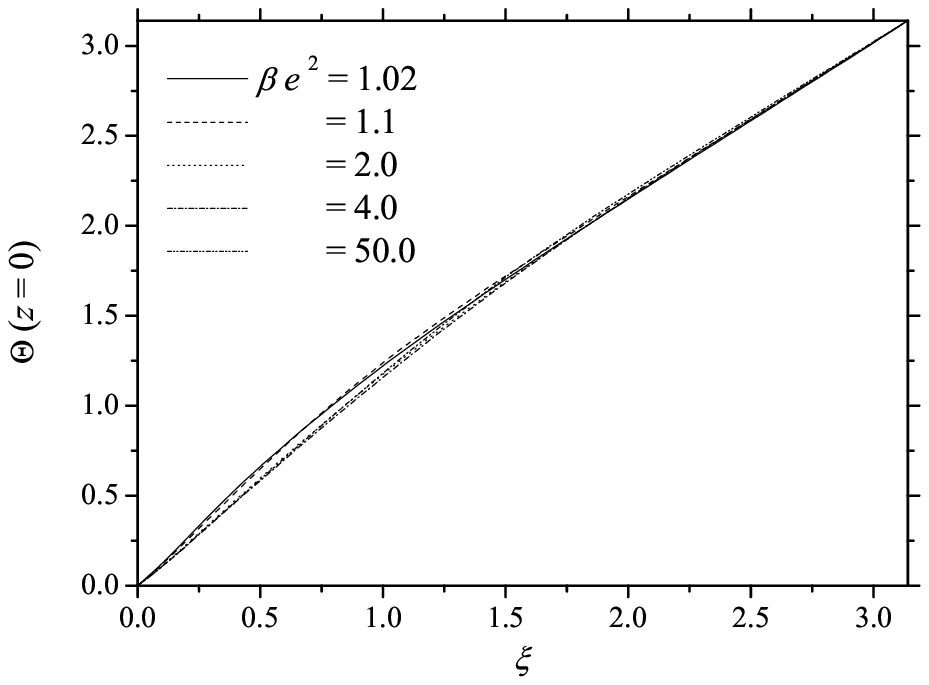}\hspace{-0.5cm} 
\includegraphics[height=7.0cm, width=9.5cm]{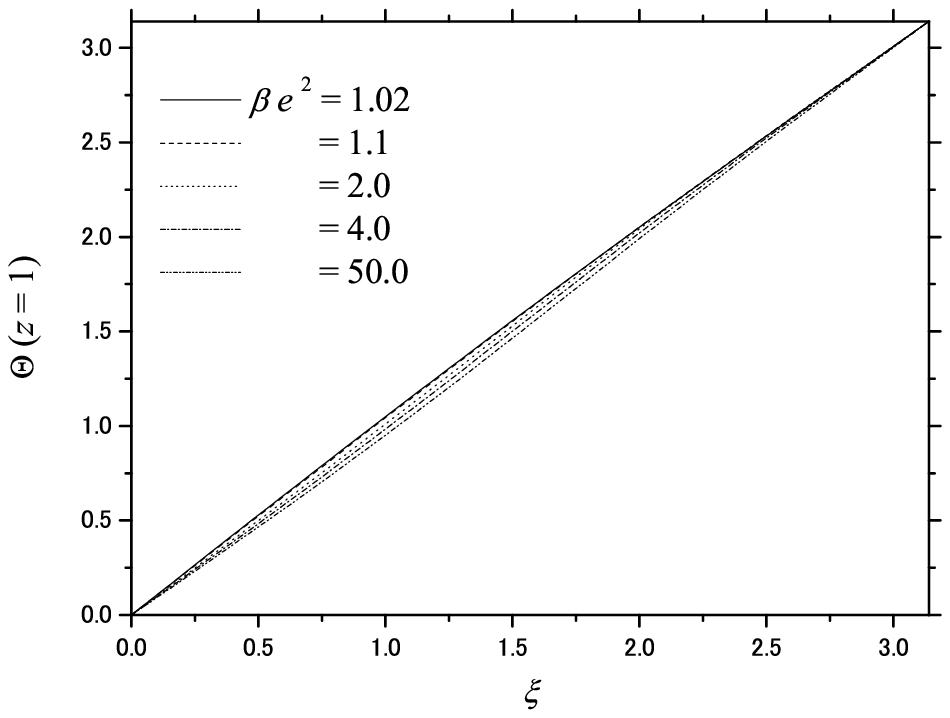}\\
{\bf (b)}~$\Theta(z,\xi)$ at $z=0$ (left) and $z=1$ (right);\\  
\caption{\label{fig4}Cuts of the ansatz functions $g(z,\xi)$ and
  $\Theta(z,\xi)$ at the boundaries for the solution $\(m,n\)=\(1,2\)$
  ($Q_H=2$) and for some values of $\beta\,e^2$ shown on the plots.}
\end{figure*}

The solutions for the ansatz functions $g\(z,\xi\)$ and
$\Theta\(z,\xi\)$ satisfying the boundary conditions \rf{bc1},
\rf{bc2}, \rf{bc3}, \rf{bc4} and \rf{bc5} are quite regular. In
Figures  \ref{fig1} and \ref{fig2} we give the plots of those
functions for the cases $(m,n)=(1,1)$ and $(m,n)=(1,2)$ respectively,
and for the values $\beta e^2=1.1$ and $5.0$ in both cases. We show
the plot in the range $0\leq \xi\leq \pi$, and the solution in the
range $-\pi\leq \xi\leq 0$, is obtained through the symmetry
\rf{symmetryeqmot}. Notice that the changes of $g\(z,\xi\)$ and
$\Theta\(z,\xi\)$ under variations of $\beta\,e^2$ are very small. In
order to visualize those changes better we also present in Figure
\ref{fig3} cuts of the
function $g$ at the borders  $\xi=0$ and $\xi=\pi$, and for several values of
$\beta\,e^2$. We also show in Figure \ref{fig4} cuts of $\Theta$ at
the borders $z=0$ and $z=1$ for several values of $\beta\,e^2$.

\begin{figure*}
\hspace{-2.5cm}
\includegraphics[height=7.5cm, width=9cm]{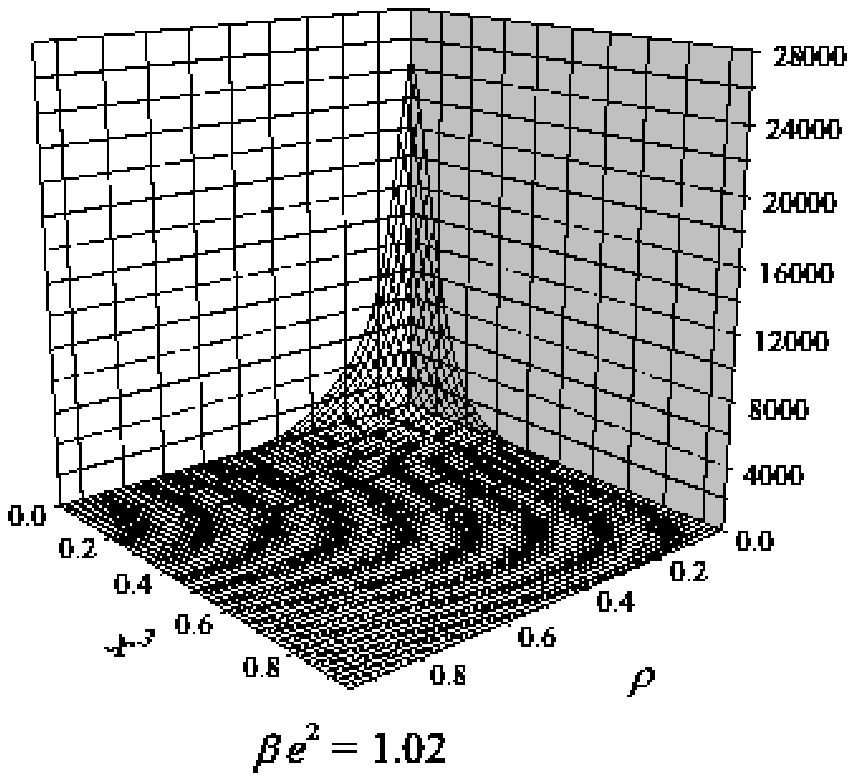}\hspace{-1cm}
\includegraphics[height=7.5cm, width=9cm]{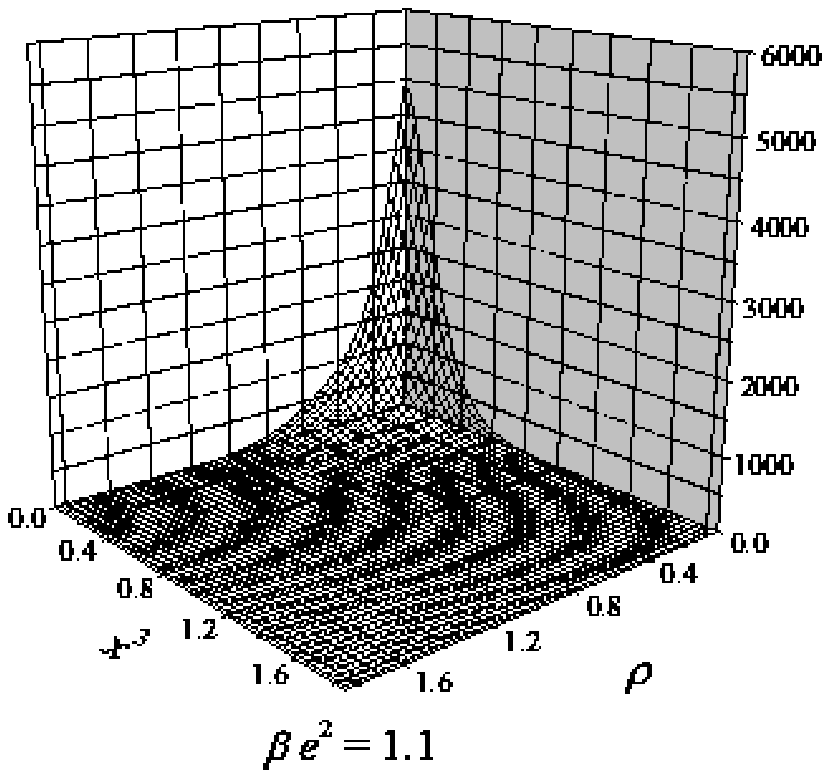}\\
\hspace{-2.5cm}
\includegraphics[height=7.5cm, width=9cm]{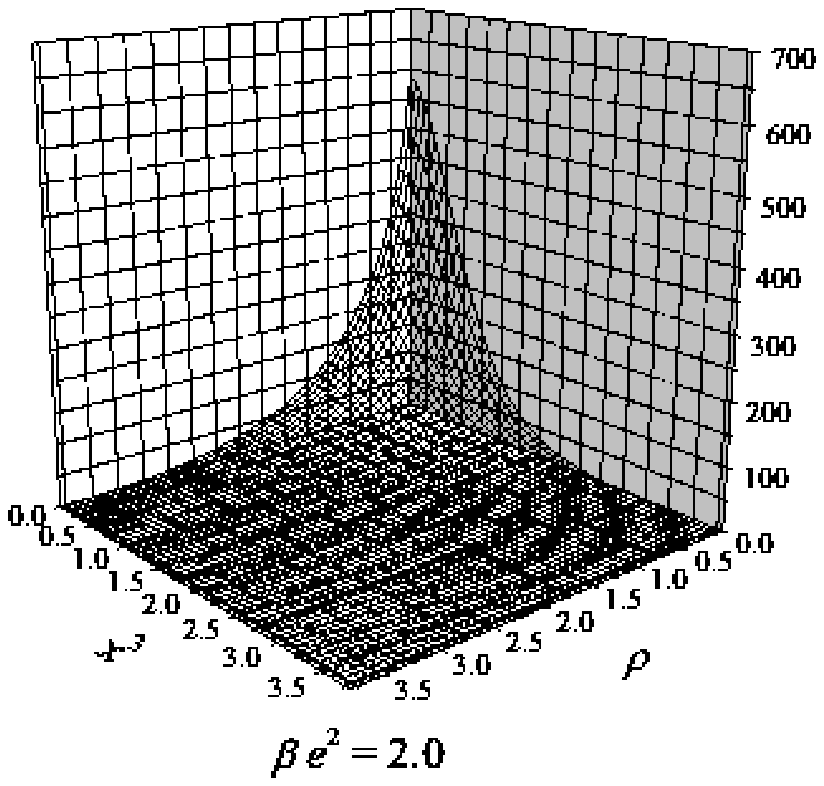}\hspace{-1cm}
\includegraphics[height=7.5cm, width=9cm]{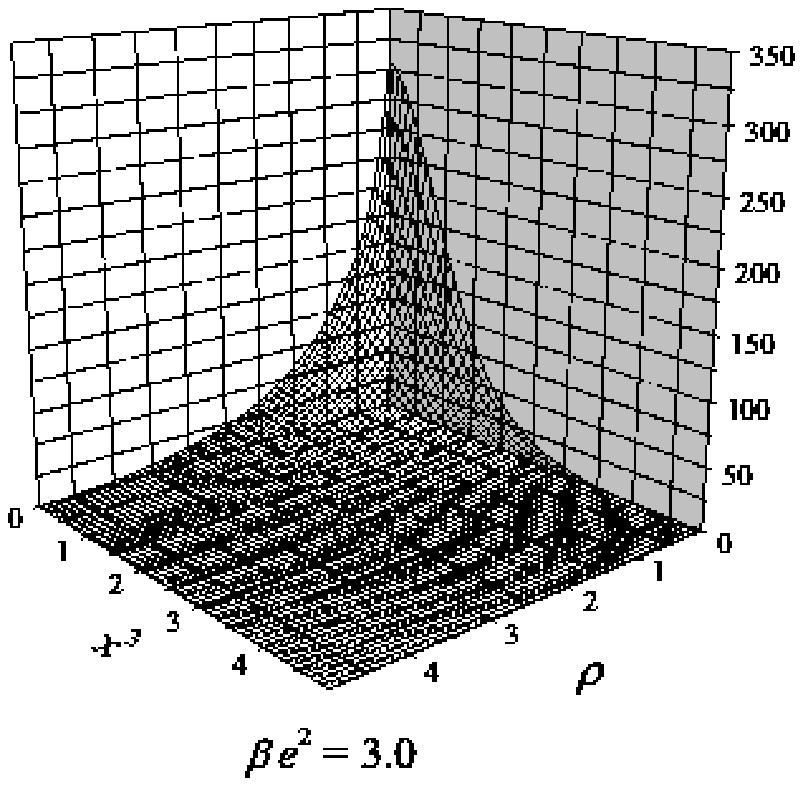}\\
\hspace{-2.5cm}
\includegraphics[height=7.5cm, width=9cm]{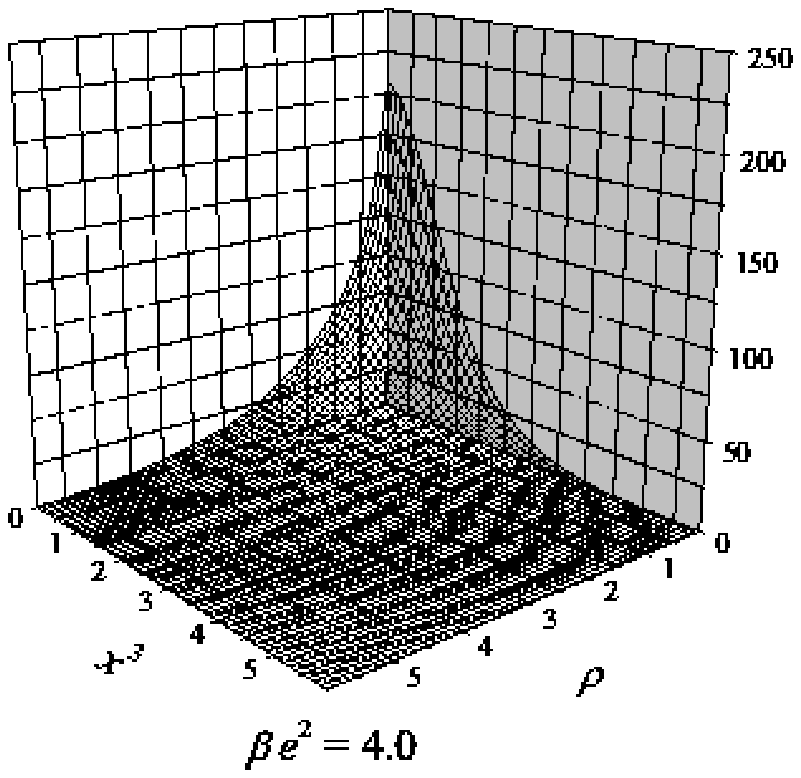}\hspace{-1cm}
\includegraphics[height=7.5cm, width=9cm]{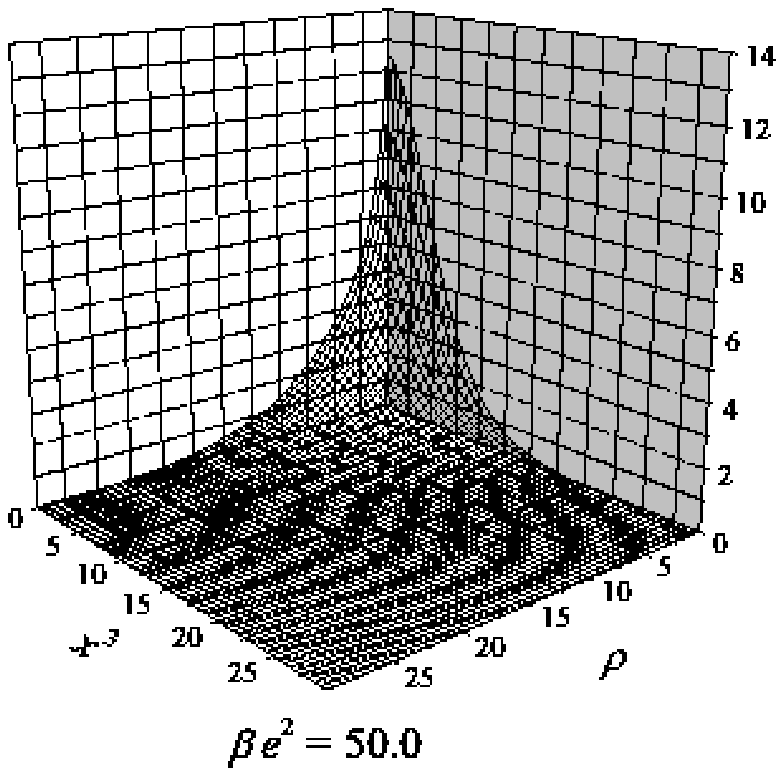}
\caption{\label{fig5}The energy density (in units of $\dfrac{M}{|e|}$)
  for the solution $\(m,n\)=\(1,1\)$  ($Q_H=1$) on the cylindrical
  coordinates plane $(\rho:=a\sqrt{z}/p,x^3)$. }
\end{figure*}

\begin{figure*}
\hspace{-2.5cm}
\includegraphics[height=7.5cm, width=9cm]{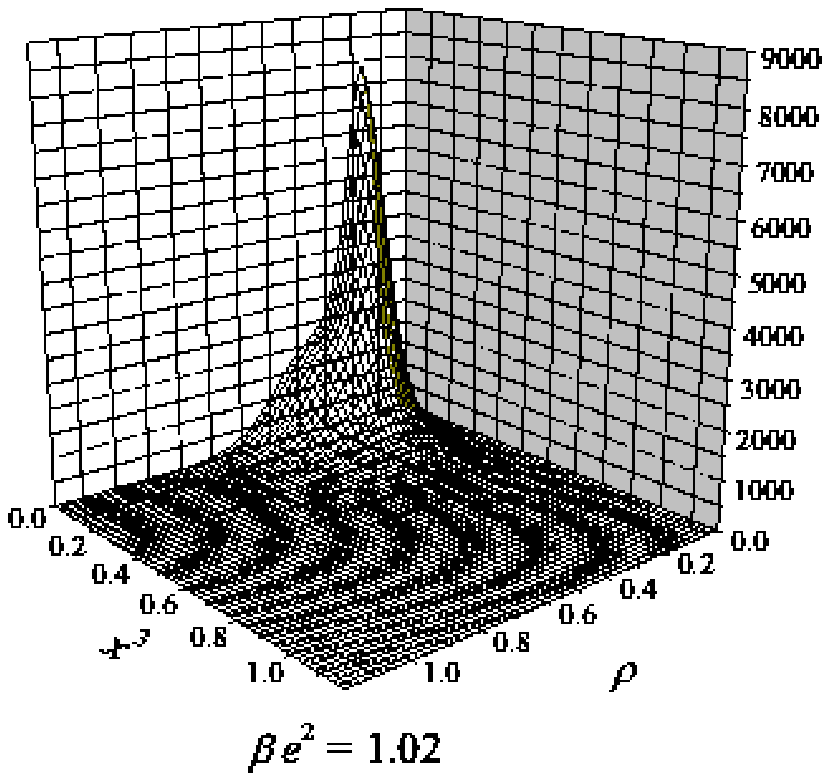}\hspace{-1cm}
\includegraphics[height=7.5cm, width=9cm]{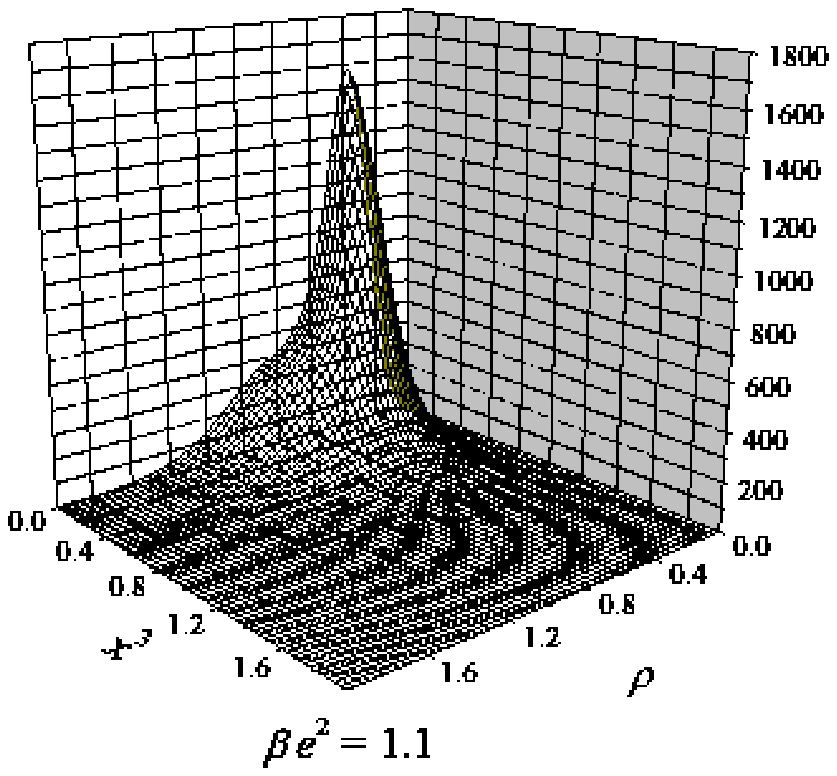}\\
\hspace{-2.5cm}
\includegraphics[height=7.5cm, width=9cm]{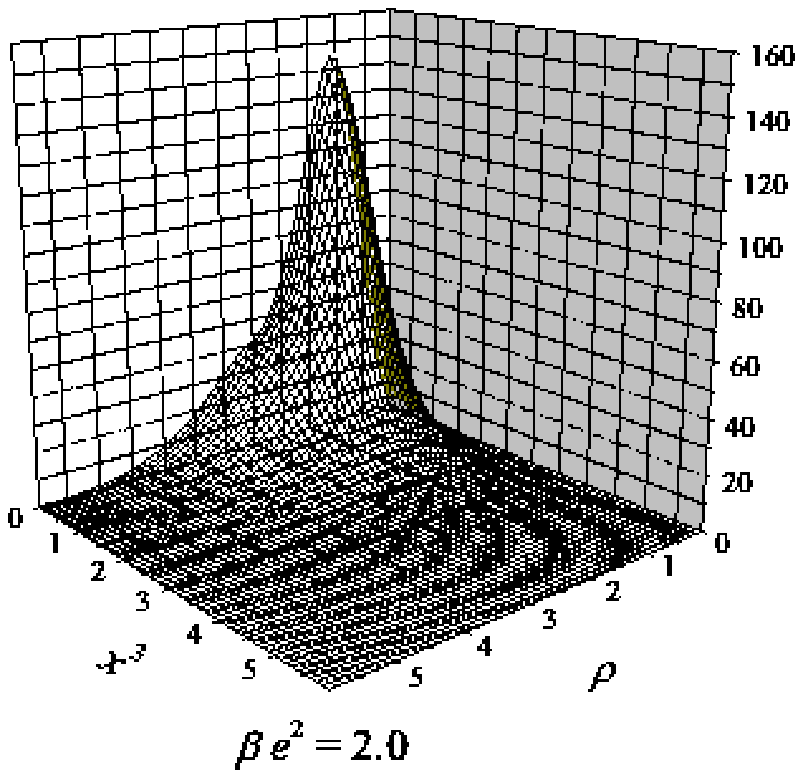}\hspace{-1cm}
\includegraphics[height=7.5cm, width=9cm]{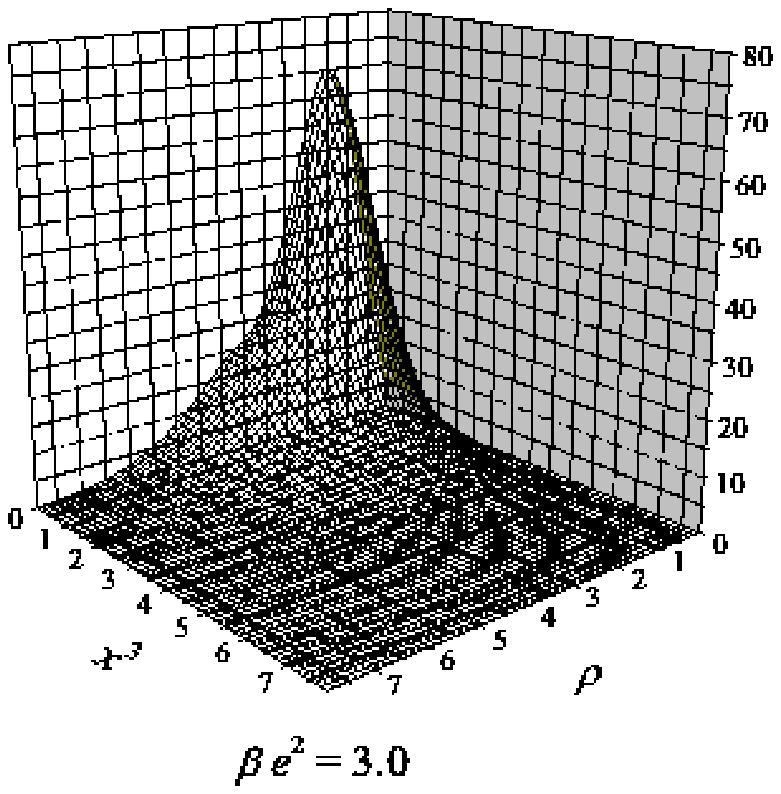}\\
\hspace{-2.5cm}
\includegraphics[height=7.5cm, width=9cm]{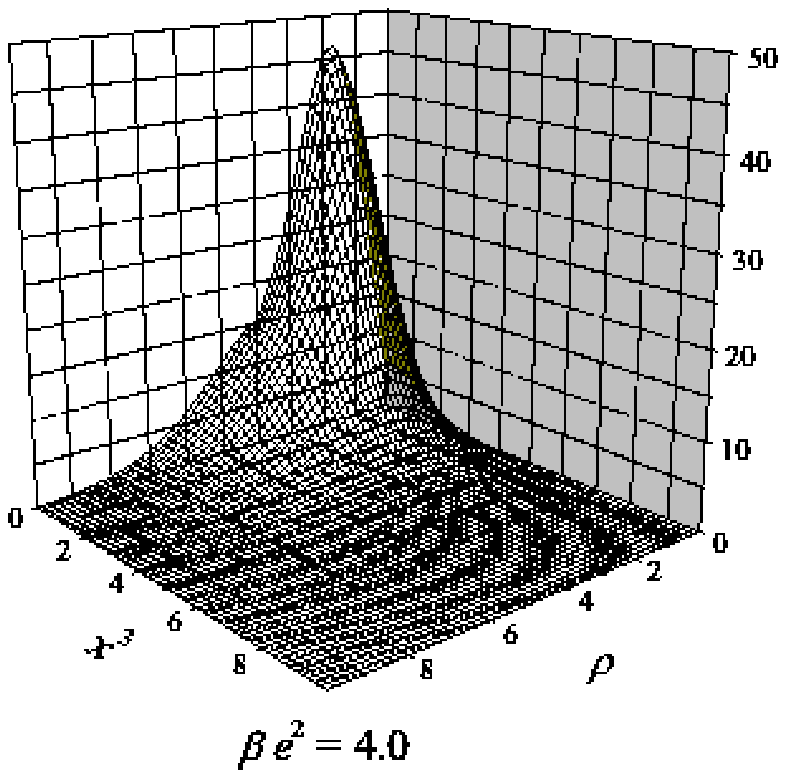}\hspace{-1cm}
\includegraphics[height=7.5cm, width=9cm]{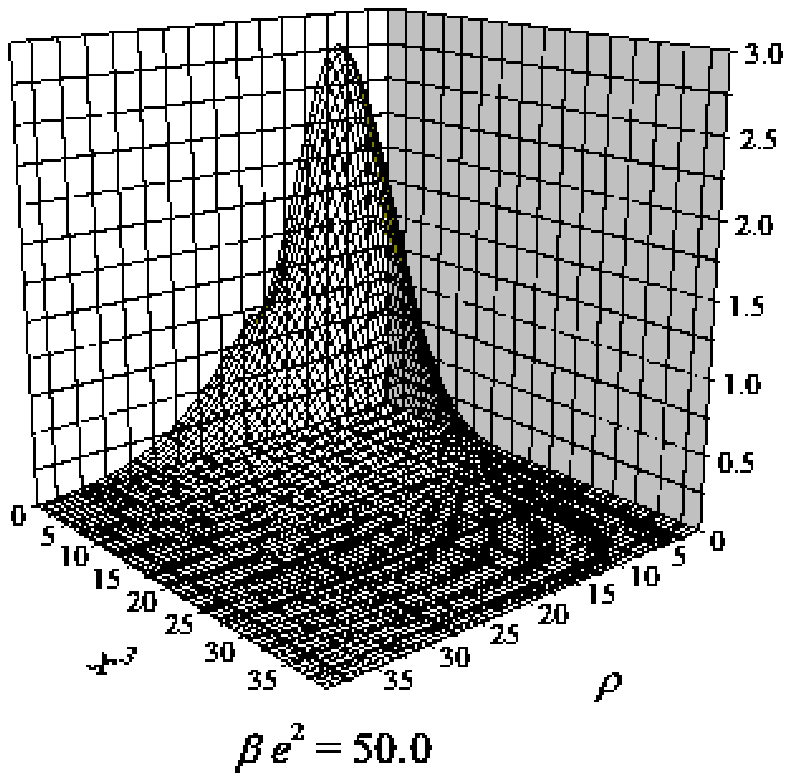}
\caption{\label{fig6}The energy density (in units of $\dfrac{M}{|e|}$)
  for the solution  $\(m,n\)=\(1,2\)$  ($Q_H=2$) on the cylindrical
  coordinates plane $(\rho:=a\sqrt{z}/p,x^3)$.}
\end{figure*}

We also present the densities of the static energy \rf{energyscaled} in the
cylindrical  
coordinate space $(\rho:=a\sqrt{z}/p,x^3)$ in Fig.\ref{fig5}  for $Q_H=1$,
and in Fig. \ref{fig6} for $Q_H=2$, for several values of $\beta e^2$. 
The following relations between two coordinates are quite useful to
visualize them: 
\begin{eqnarray}
\xi=\pm
\cos^{-1}\Bigl[\frac{a^2-\rho^2-(x^3)^2}{a^4-2a^2(\rho^2-(x^3)^2)+(\rho^2+(x^3)^2)^2}\Bigr],~~   
z=\frac{4a^2 \rho^2}{(a^2+\rho^2+(x^3)^2)^2}\,. 
\end{eqnarray}
The energy density of $\(m,n\)=\(1,1\)$ ($Q_H=1$) solution has the lump
shaped, and of the $\(m,n\)=\(1,2\)$ ($Q_H=2$) solution 
exhibits the toroidal configuration. For both case, one easily observes
the shrinking of the solutions as $\beta e^2\to 1$, and so an increase
of the density around the origin.  

\begin{figure*}
\hspace{-2.5cm}
\includegraphics[height=7.5cm, width=9cm]{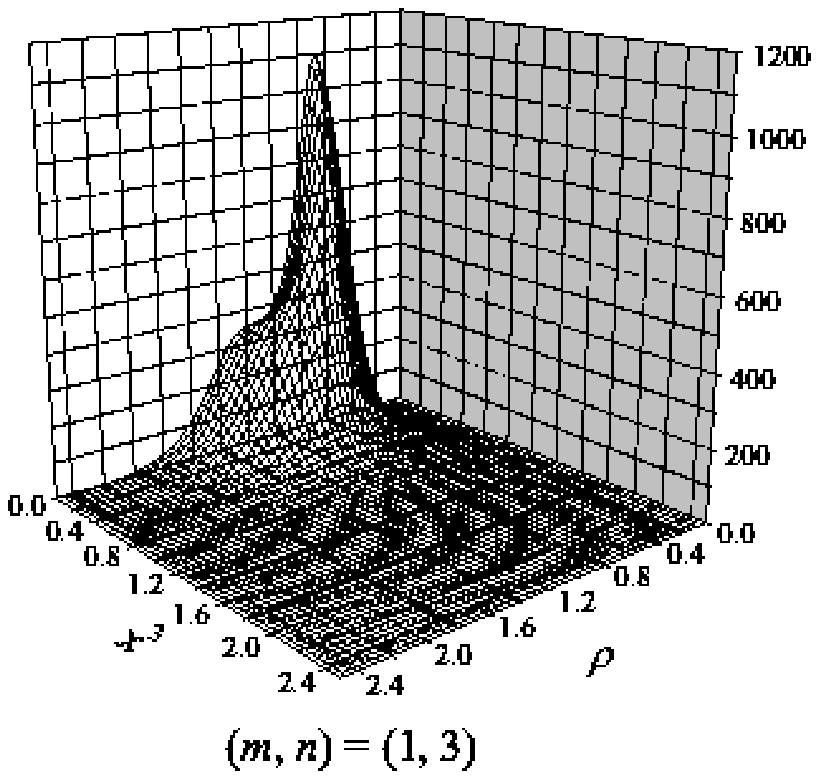}\hspace{-1.0cm}
\includegraphics[height=7.5cm, width=9cm]{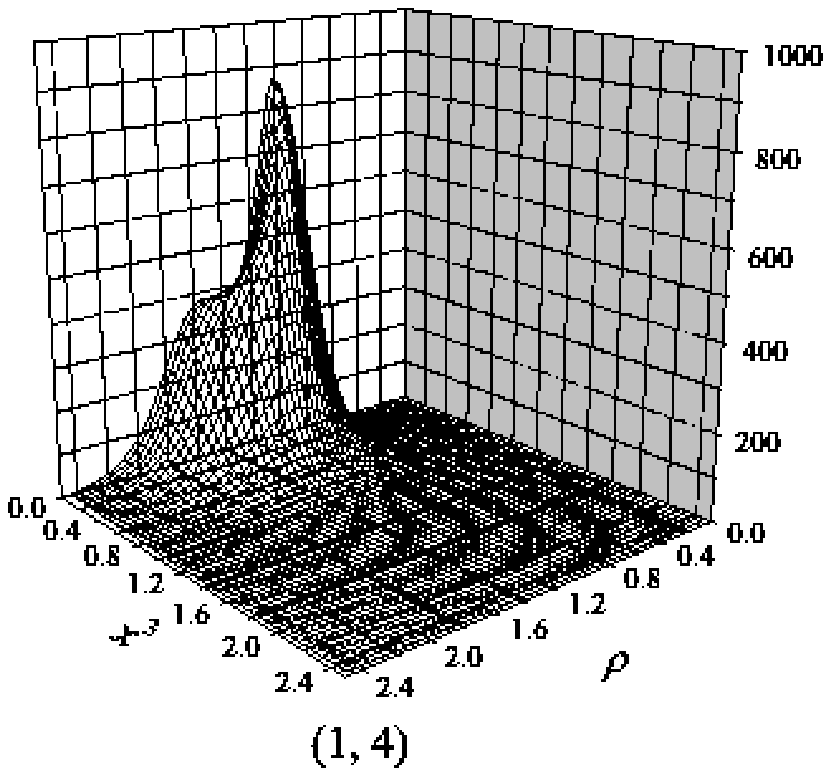}\\
\hspace{-2.5cm}
\includegraphics[height=7.5cm, width=9cm]{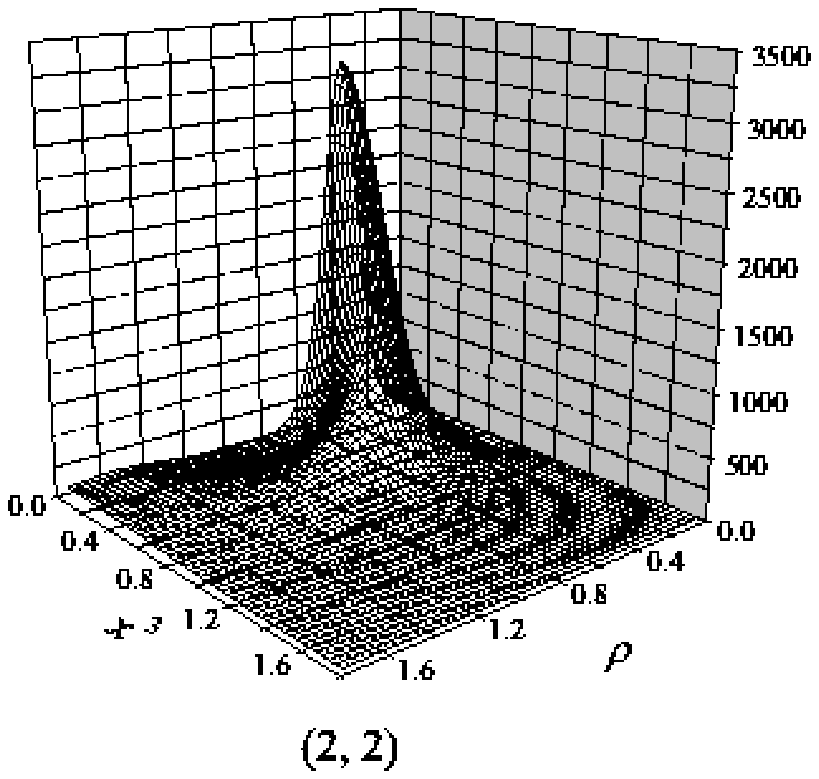}\hspace{-1.0cm}
\includegraphics[height=7.5cm, width=9cm]{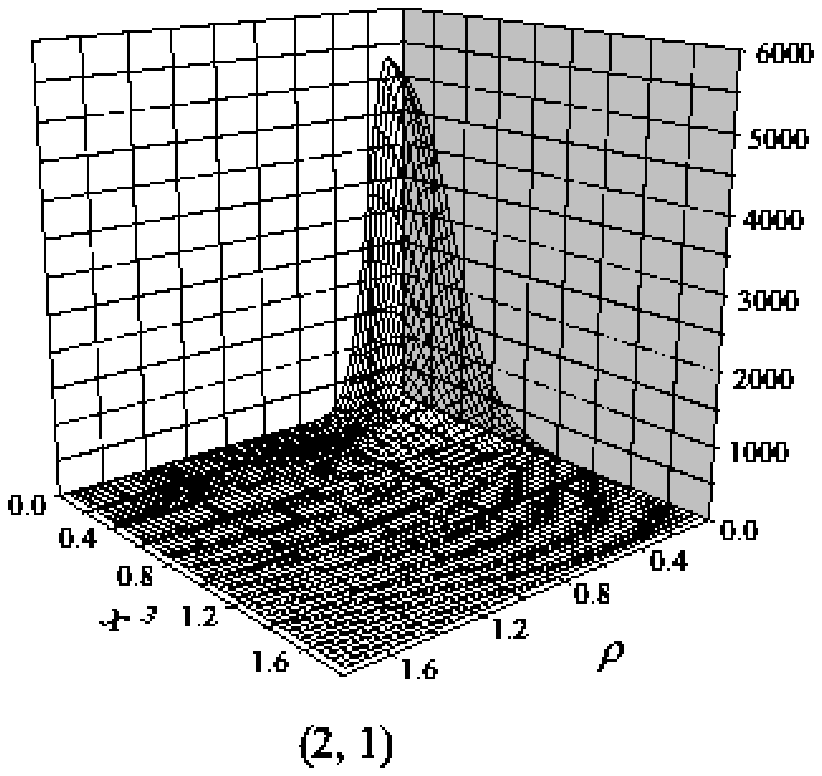}\\
\hspace{-2.5cm}
\includegraphics[height=7.5cm, width=9cm]{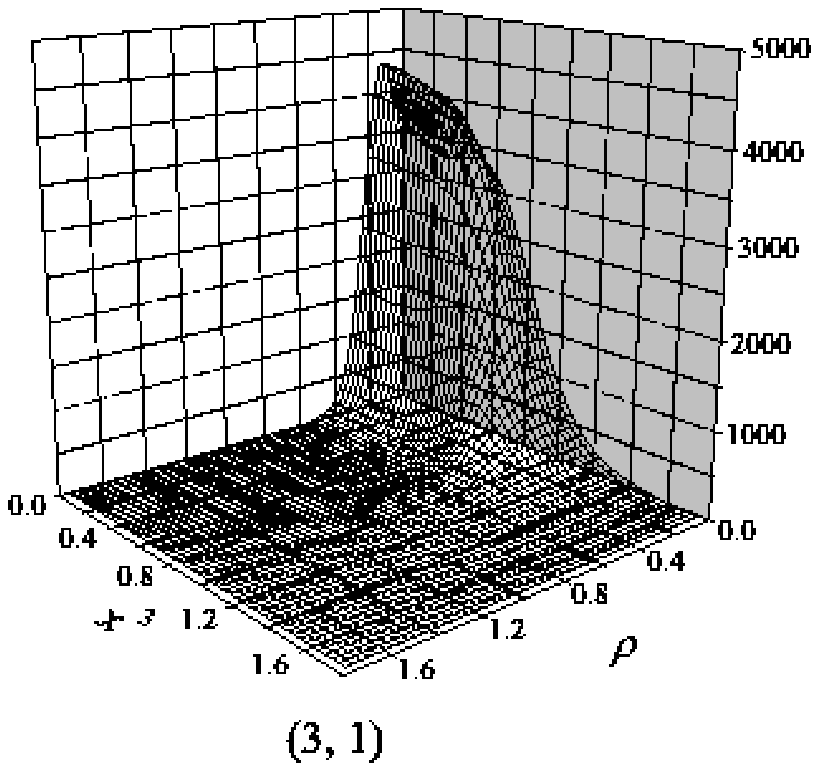}\hspace{-1.0cm}
\includegraphics[height=7.5cm, width=9cm]{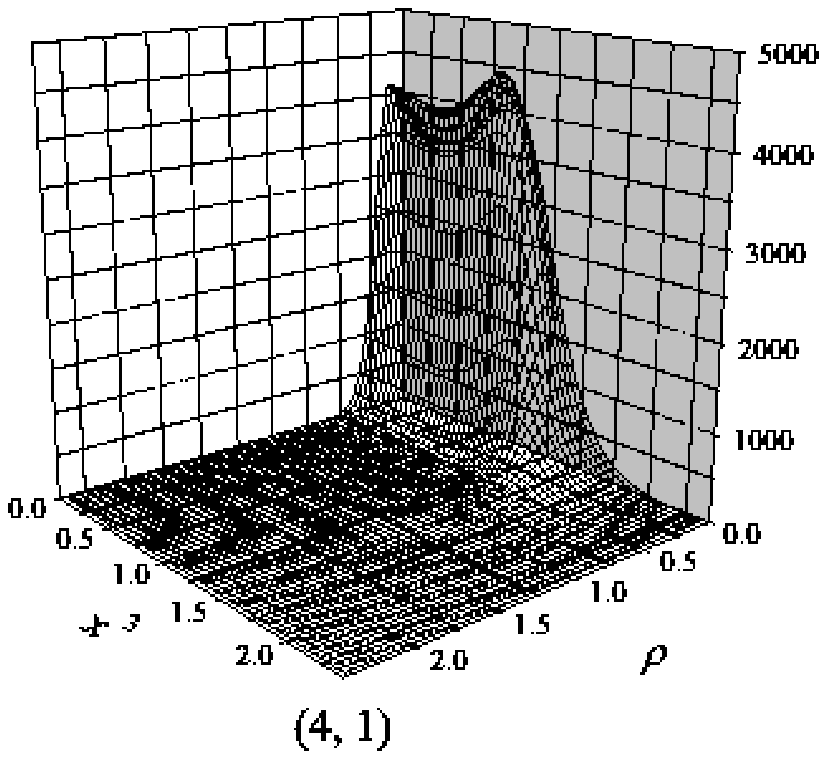}
\caption{\label{fig7} The energy density (in units of $\dfrac{M}{|e|}$)
  for the solutions with  
higher Hopf charges $Q_H>2$ for $\beta e^2=1.1$, on the cylindrical
  coordinates plane $(\rho:=a\sqrt{z}/p,x^3)$.}
\end{figure*}

In Fig.\ref{fig7}, we display the energy densities for solutions with
charges $Q_H>2$.  
The $n=3,4$ solitons exhibit the toroidal shape. The radius of the
tori increases as $n$ grows.  
On the other hand, all $n=1$ solitons seem to be lump shaped,
however, for $m\geqq 2$ they have a depression  close to   
the origin. The size grows especially in the $x^3$ direction as the
charge increases,  and as that happens a second peak   
gradually  emerges. 

\vspace{.5cm}

\noindent {\bf Acknowledgments} L.A.F. is grateful for the
hospitality at the Department of Physics of the Tokyo University of
Science, and the Department of Mathematical Physics of the Toyama Prefectural
University, where this work was initiated. 
N.S. and K. T. would like to thank  the kind hospitality 
at Instituto de F\'isica de S\~ao Carlos, Universidade de S\~ao Paulo.
The authors acknowledge the financial support of 
FAPESP (Brazil). L.A.F. is partially supported by CNPq.

\newpage

\newpage

\end{document}